\documentclass[preprint,pre,showpacs,preprintnumbers,amsmath,amssymb]{revtex4}

\usepackage{latexsym}
\usepackage{amsmath}
\usepackage{amsfonts} 
\usepackage{epsfig}
\usepackage{graphicx}
\usepackage{dcolumn}
\usepackage{bm}

\newcommand{\Vektor}[1]{\mbox{\boldmath $#1$}}
\newcommand{\tdiff}[2]{\frac{d #1}{d #2}}

\newcommand{\pdiff}[2]{\frac{\partial #1}{\partial #2}}

\newcommand{\abs}[1]{\left| #1 \right|}

\newcommand{\bea}{\begin{eqnarray}}
\newcommand{\eea}{\end{eqnarray}}
\newcommand{\beann}{\begin{eqnarray*}}
\newcommand{\eeann}{\end{eqnarray*}}

\newcommand{\order}[1]{{\mathcal O}\left( #1 \right)}

\newcommand{\nn}{\nonumber}

\begin{document} 

\preprint{q-bio.TO/0407029}

\title{Multicellular Tumor Spheroid in an off-lattice Voronoi/Delaunay cell model}
\author{Gernot Schaller}
\affiliation{Institut f\"ur Theoretische Physik,
        Technische  Universit\"at Dresden,
        D-01062  Dresden, Germany}
\email{{schaller@theory.phy.tu-dresden.de}}
\author{Michael Meyer-Hermann}
\affiliation{Centre for Mathematical Biology, 
        Mathematical Institute, 24-29 St. Giles',
        Oxford University, Oxford OX1 3LB, United Kingdom}
\date{\today}

\begin{abstract}
We study multicellular tumor spheroids by introducing a new three-dimensional agent-based Voronoi/Delaunay
hybrid model. In this model, the cell shape varies from spherical in thin solution to convex polyhedral in 
dense tissues. The next neighbors of the cells are provided by a weighted Delaunay triangulation with in average
linear computational complexity. 
The cellular interactions include direct elastic forces and cell-cell as well as cell-matrix adhesion.
The spatiotemporal distribution of two nutrients -- oxygen and glucose
-- is described by reaction-diffusion equations.
Viable cells consume the nutrients, which are converted into biomass
by increasing the cell size during $\rm G_1$-phase.

We test hypotheses on the functional dependence of the uptake rates and use the computer simulation to find suitable
mechanisms for induction of necrosis.
This is done by comparing the outcome with experimental growth curves,
where the best fit leads to an unexpected ratio of oxygen and glucose uptake rates.
The model relies on physical quantities and can easily be generalized
towards tissues involving different cell types. In addition, it
provides many features that can be directly compared with the experiment.
\end{abstract}    

\pacs{  45.05.+x,
        82.30.-b,
        87.*,
        02.70.-c
        }

\keywords{agent-based mathematical model, Voronoi/Delaunay, multicellular tumor spheroid,
tumor growth, cell nutrient consumption}

\maketitle


\section{Introduction}\label{Sintroduction}

The spatiotemporal dynamics of individual cells often leads to the emergence of fascinating complex patterns 
in cellular tissues.
For example, during embryogenesis it is hypothesized that these complex patterns develop with the aid of mechanisms such as 
diffusing messengers and cell-cell contact.
Sometimes these patterns can be described very well with a simple model. 
Such mathematical models can help to test hypotheses in {\em in silico} experiments thereby circumventing real experiments
which are very often expensive and time-consuming.
However, since the local nature of cell-cell interactions is not precisely known one is often restricted to 
compare the global outcome following from different hypotheses with
experimental data.
Unfortunately, there are -- unlike in theoretical physics -- no established first principle theories in cell tissue modeling 
which explains that there is a variety of models on the market, which can be classified as follows:

Firstly, there is a class of models where one derives continuum equations for the cell populations. 
In analogy to many-particle physics one replaces the actual information on every cell by a cellular density. Consequently, the equations of motion can
be simplified considerably to a differential equation describing the spatiotemporal dynamics of a cell type.
In practice these equations do very often have the type of reaction-diffusion equations \cite{murray2002a}. 
The volume-integral of such equations results in the global dynamics of a whole population (e.g. predator-prey-models), 
where only the temporal development of the total population is monitored.
Note however, that cellular interactions can only be modeled effectively with these approaches.
Also, the discrete and individual nature of cells is completely neglected. 

The discrete nature can be taken into account by deriving master equations for the population number
on every volume element \cite{shnerb2000}. By mapping these master equations to a Schr\"odinger equation one is able to identify an
Hamilton operator that allows a physicist to apply the mathematical framework of quantum field theory to systems such as
cell tissues.
For example, in the simple case of Lotka-Volterra equations
\cite{murray2002a} this method leads to mean-field equations that
resemble the Lotka-Volterra equations.
The renormalized numerical results \cite{bettelheim2001} however may
disagree qualitatively with the mean-field approximations.
Consequently, the discrete nature of cells may not always be neglected.
Still, the above quantization assumes all agents to be identical and indistinguishable and inevitably neglects the individuality of cells. 
Therefore, features such as cell shape and differences in cell size
or internal properties are not considered in this class of models.

This is different in the third class of agent-based models, where cells are represented by individually interacting objects.
Since now every single cell must be included in the computer simulations the computational intensity increases considerably. 
This however opens often the possibility to choose the interaction rules intuitively from existing observations.
These models are usually restricted to a certain cell shape, which enables one to sub-classify them further:
In lattice-based models \cite{deutsch1993,drasdo2002} the cellular shape is usually already defined by the shape of the elementary cell of the lattice, 
such as e.g. cubic \cite{meyerhermann2002} or hexagonal \cite{dormann2002,beyer2002}. 
Off-lattice models usually restrict to one special cell form and consider
slight perturbations (e.g. deformable spheres \cite{drasdo1995,galle2005a} or deformable ellipsoids \cite{palsson2000,palsson2001,dallon2004}). 
In other off-lattice models the geometrical Voronoi tessellation \cite{honda2000,schaller2004} is used, 
which allows for more variations in cell shape and size. 
In addition, it comes very close to the polyhedral shape observed for some cell types \cite{honda1978}.
An important advantage of off-lattice models is that perturbations from the inert cell shape can give rise to physically well-defined
cellular interaction forces, whereas in lattice-based models one is usually forced to introduce effective interaction rules which makes it difficult
to relate the model parameters to experimentally accessible quantities.

Since cell shape and function are usually closely connected 
(e.~g.~fibroblasts in the human skin do have a different shape 
than melanocytes or keratinocytes), there are some models
that try to reproduce any possible cell shape. 
For example in the extended Potts model \cite{graner1992,savill2003,stott1999,turner2002} 
one has spins on several lattice nodes describing a single cell. 
The dynamics of these spins is calculated by minimizing an energy
functional. The often-used Metropolis algorithm
tests several spin flips for a decrease of the energy. 
A Metropolis time step is defined as having performed as many checks for
spin flips as there are spins.
The parameters in the energy functional have to be determined heuristically
as it is difficult to map them
to experimentally accessible microscopic properties.
For example, volume conservation is usually handled by a penalty term
which acts equally strong for both compression and elongation. 
The usual practice of relating the Monte-Carlo time step to physical time
is not unique:
There are cellular proliferation times, cellular
compression relaxation times etc.
Finally, the enormous number of spins required to appropriately describe a single
cell leads to an enormous computational complexity that restricts the
model to small cell numbers.
This problem is circumvented in force-based models. For example
in \cite{meyer_hermann2004} the relation of cell shape and cell
motility has been investigated in a model that represents cells as a
collection of cell fragments on a lattice.
Other models describe cell shape on a 2-dimensional hypersurface by a changing
number of polygonal nodes \cite{weliky1990,weliky1991}, which is also
computationally expensive.
In \cite{honda2004}, the initial configuration of the nodes bordering
the polyhedral cells is deduced from a Voronoi tessellation of the
cell centers, whereas the Voronoi concept is discarded during the
dynamics, since every border node has its own dynamics. 
Generally, the latter models always need a large number of general coordinates to
define the shape or status of a cell and are therefore restricted to a
relatively small number of cells -- even at present computational power.

Balancing these reasons in the context of the aimed description of 
{\em in vitro} tumor growth data we decided to use an off-lattice agent-based
model, where one has the advantage of allowing continuous cell positions. 
Therefore the extent by which cellular interactions have to be replaced by effective automaton rules is much smaller than in corresponding 
cellular automata \cite{dormann2002}. 
In addition, the model parameters can be directly measured in independent experiments. 
The enormous computational intensity common to most existing off-lattice models is due to two effects:
Firstly, some off-lattice models \cite{drasdo2002} use effective stochastic interaction rules, 
which require stochastic solution methods such as the Metropolis algorithm. 
The infinite number of possibilities in a continuous model however requires a large part of the phase space to be tested.
Secondly, the determination of the neighborship topology for local interactions requires sophisticated algorithms.
Our model uses the weighted Delaunay triangulation which provides the correct neighborship topology for a set of
spheres with different radii with in average constant access \cite{schaller2004}.
In addition, the model is dominantly deterministic which abolishes the necessity to test irrelevant parts of the phase space.

Unlike in two dimensions, where tumor cells in {\em in vitro} setups will proliferate without limitation, there exist growth limitations
on tumor cell populations forming solid spheroidal cell aggregates in three dimensions \cite{folkman1973}.
This limitation of growth is presumably due to both contact inhibition -- which is also active in two dimensions \cite{galle2005a} -- and
nutrient depletion in the interior of the spheroid.
Initially, the cell number grows exponentially and enters a polynomial growth phase after some days in culture.
Finally, a saturation of growth is observed for many spheroid systems \cite{freyer1986a}.
The final stages of spheroid growth exhibit a typical pattern in the cross-sections:
An internal necrotic core is surrounded by a layer of quiescent cells -- which do not proliferate -- and on the 
outside one has a layer of proliferating cells \cite{mueller_klieser2002}.
The final stage depends critically on the supply with nutrients such as oxygen and glucose.
The model we have implemented enables us to model $\order{10^5}$ cells which is in agreement with cell numbers
observed in multicellular tumor spheroid systems \cite{freyer1986a}.
We will demonstrate that the growth curves measured in \cite{freyer1986a} 
for different nutrient concentrations can be reproduced using a single parameter set and simple assumptions for cellular interactions.


\section{The cell model}\label{Scellmodel}

In our model we assume cells to be deformable spheres with dynamic radii, which is motivated by the experimental observation
that cells in a solution tend to be spherical -- presumably in order to minimize their surface energy.
Consequently, we treat all deviations from this spherical form as
perturbations from the inert cellular shape.

The model is agent-based (sometimes also called individual-based), i.~e.~every biological cell is represented 
by an individual object.
These objects interact locally with their next neighbors 
(those that follow from the weighted Delaunay triangulation)
and with a reaction-diffusion grid (for nutrients or growth signals).
Each cell is characterized by several individual parameters such as
position, a radius, the type corresponding to biological classifications,
the status (position in the cell cycle),
cellular tension, receptor and ligand concentrations on the cell membrane,
an internal clock, and cell-type specific coupling constants for elastic and adhesive interactions.
Since we assume the inert cell shape to be spherical,
the power-weighted Delaunay triangulation \cite{schaller2004}
is a perfect tool to determine the neighborship topology. 

\begin{figure*}
\begin{tabular}{cc}
\begin{minipage}{0.47\linewidth}
\includegraphics[height=4cm]{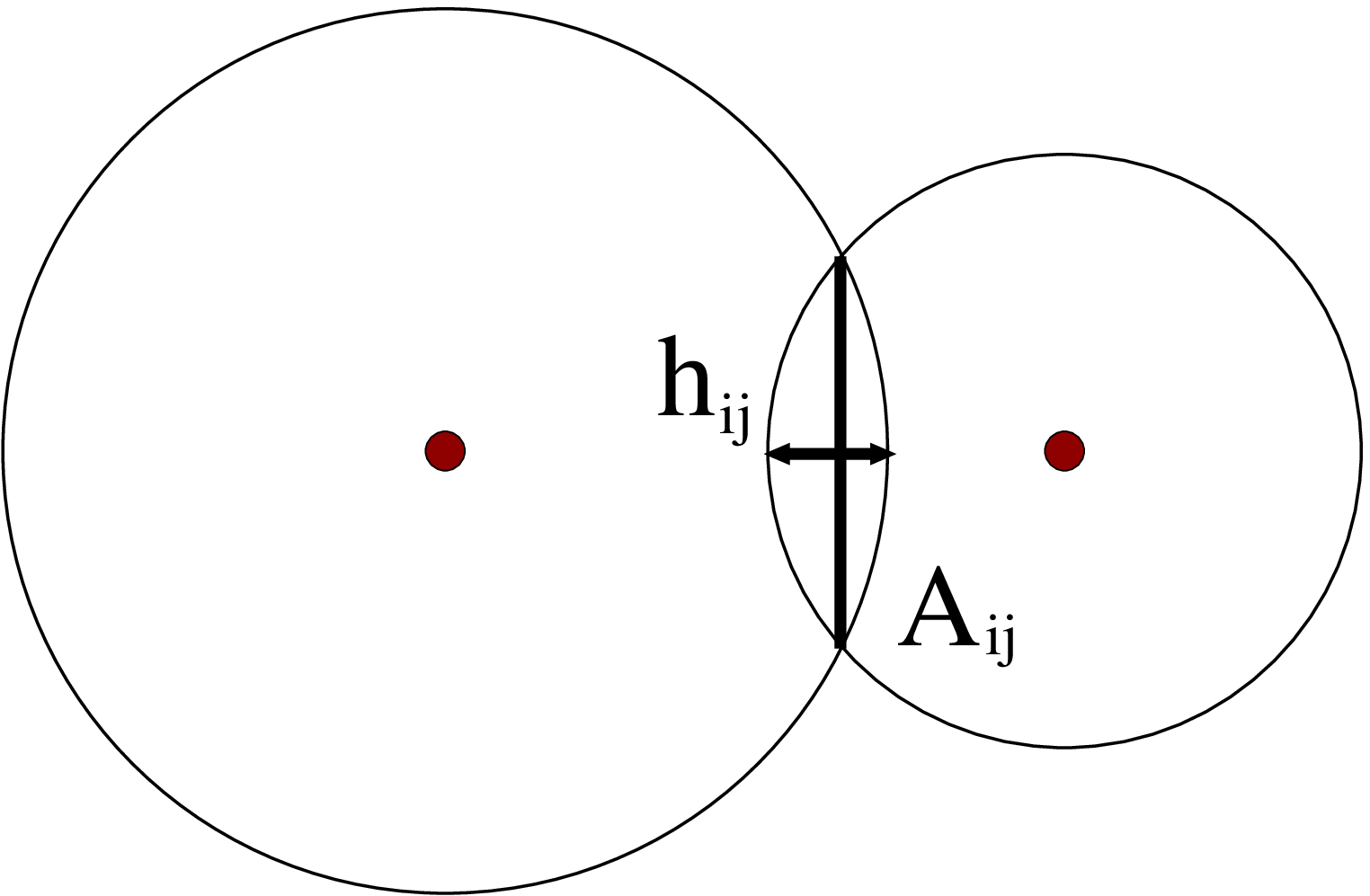}
\end{minipage}
&
\begin{minipage}{0.47\linewidth}
\includegraphics[height=4cm]{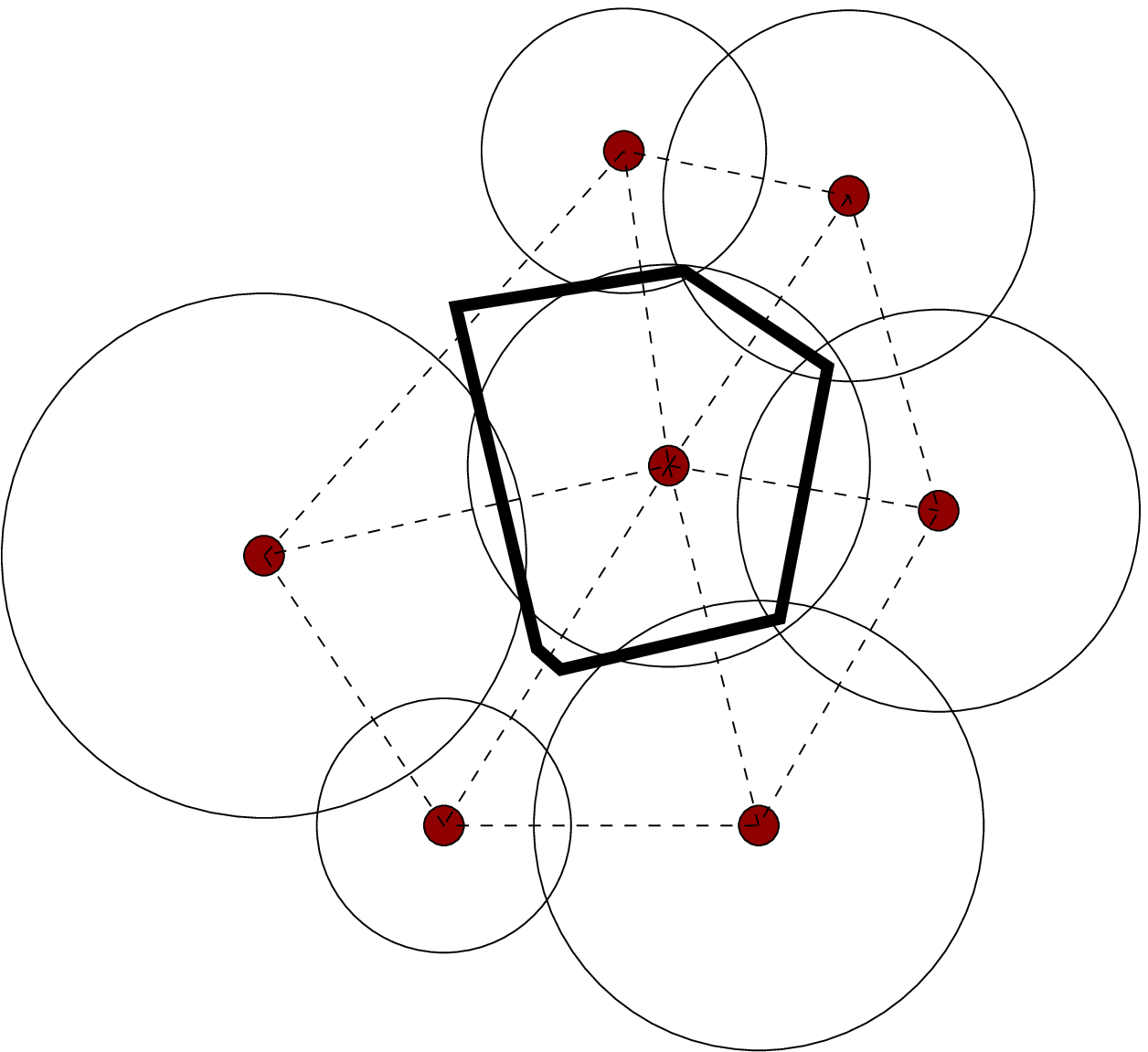}
\end{minipage}\\
\begin{minipage}{0.45\linewidth}
\caption{\label{Fsphere_overlap}Two-dimensional illustration of
inter-penetrating spheres with maximum overlap $h_{ij}$ and sphere
contact surface $A_{ij}$ (marked bold). 
In reality, the spheres will deform and generate a repulsive force.}
\end{minipage}
&
\begin{minipage}{0.45\linewidth}
\caption{\label{Fvoronoi_correction}Within dense tissues,
many-sphere-overlaps can occur. If in this case the Voronoi contact
surface (marked with a bold line) is smaller than the sphere contact
surface, it will provide a more realistic estimate of the cellular contact surfaces.}
\end{minipage}

\end{tabular}
\end{figure*}


\subsection{Elastic and adhesive Cell-Cell interactions}

Following a model of Hertz \cite{hertz1882,landau1959} -- which has
already been used in the framework of cell tissues
\cite{galle2005a,wei2001} -- the absolute value of the elastic force between two spheres with radii $R_i$ and $R_j$ 
can, for small deformations, be described as
\bea\label{Ehertz}
F_{ij}^{\rm el}(t) = \frac{h_{ij}^{3/2}(t)}{\frac{3}{4}\left(\frac{1-\nu_i^2}{E_i} + \frac{1-\nu_j^2}{E_j}\right)\sqrt{\frac{1}{R_i(t)} + \frac{1}{R_j(t)}}}\,,
\eea    
where $E_{i/j}$ and $\nu_{i/j}$ represent the elasticities and Poisson ratios of the spheres, respectively.
The quantity $h_{ij}=\max\{0, R_i + R_j - \abs{\Vektor{r}_i - \Vektor{r}_j}\}$ 
represents the maximum overlap the spheres would have if they would not deform but interpenetrate each other,
see figure \ref{Fsphere_overlap}. 
In principle the repulsive force resulting from
(\ref{Ehertz}) could be overturned since it does not diverge for large
overlaps. However, additional
mechanisms (contact inhibition) insure that in practice the cells
will respect a minimum distance from each other. In addition, the
overlaps lead to a deviation of the actual cell volume (set
intersection of Voronoi and sphere volume) from the intrinsic (target)
cell volume. Therefore the cell volume is only approximately conserved
within this approach.

In reality this model might not be adequate for cells:
Firstly, the mechanics of the cytoskeleton is not well represented
which might yield other than purely elastic responses (see
e.~g.~\cite{alt1995,dallon2004}).
Secondly, equation (\ref{Ehertz}) represents only a first order
approximation which is valid for small virtual overlaps $h_{ij} \ll
\min\{R_i, R_j\}$ only. As cellular mechanics is known to be not
only viscoelastic but also viscoplastic \cite{verdier2003}, 
a more exact approach would follow \cite{palsson2000,palsson2001} by
replacing cells by equivalent
networks containing elastic and viscous (internal cell friction)
elements. However, the parameters required for such a model should
either be measured for every cell type independently or
they should be derived from a microscopic model of the cytoskeleton
such as e.g. tensegrity structures \cite{ingber2003a,ingber2003b}, 
which is beyond the scope of this article.
Consequently, internal cell friction is neglected.
In addition, the Hertz model is only valid for two-body
contacts, since for an exact treatment prestress and the difficult
elastic problem of multiple overlaps will have to be
considered as well. Therefore, especially in the case of multiple
sphere overlaps (cf. figure \ref{Fvoronoi_correction}) the Hertz model
will underestimate the actual repulsion.
 
However, in this article we would like to restrict to the simple purely
elastic model (\ref{Ehertz}), since it allows the independently measurable 
experimental quantities $\nu_i$ and $E_i$ to be directly included.

Intercellular adhesion in a tissue is mediated by receptor and ligand molecules that are distributed on the cell membranes. 
For simplicity, we neglect a possible dynamical clustering of adhesion molecules and
assume them to be -- in average -- uniformly distributed. The
resulting average adhesive forces
between two cells should then scale with their
contact area $A_{ij}$ (see also e.~g.~\cite{palsson2000}) and can be estimated as
\bea\label{Ereclig1}
F_{ij}^{\rm ad} = A_{ij} f^{\rm ad} \frac{1}{2} \left(c_i^{\rm rec} c_j^{\rm lig} + c_i^{\rm lig} c_j^{\rm rec}\right)\,,
\eea
where the receptor and ligand concentrations $c_i^{\rm rec/lig}$ are
assumed to be normalized (i.~e.~$0 \le c_i^{\rm rec/lig} \le c_i^{\rm
rec/lig:max} \le 1$) 
without loss of generality, since the -- globally valid -- coupling
constant $f^{\rm ad}$ can always be rescaled by absorbing the maximum
possible densities of receptors and ligands. Therefore the receptor
and ligand concentrations do not have units but just represent the
binding strength relative to a maximum binding absorbed in $f^{\rm ad}$ 
within this model.
The contact surface area $A_{ij}$ can be estimated using the contact
surface of two overlapping spheres $A_{ij}^{\rm sphere}$ -- see figure
\ref{Fsphere_overlap}. 

Two issues need to be discussed in this respect: 
Firstly, the Hertz model predicts a contact surface of 
$A_{ij}^{\rm Hertz} = \pi (h_i + h_j) R_i R_j/(R_i + R_j)$, which is
in the physiologic regime of parameters considerably smaller
than the spherical contact surface 
$A_{ij}^{\rm sphere} = \pi (h_i R_i + h_j R_j - h_i^2/2 -
h_j^2/2)$. 
However, the spherical contact surfaces describe real
tissue much more realistically than the Hertz contact surface, which
should consequently rather be termed effective in the context of
cellular interactions.
In the used physiologic regime of overlaps the two contact surfaces
have the same scaling in the first order. Therefore, a rescaling of
the effective adhesive constant $f^{\rm ad}$ will replace the spherical contact
surface by the Hertz contact surface.
Secondly, in dense tissues the spherical contact surface is not a valid
description anymore, since the contact surfaces of many spheres might
overlap as in figure \ref{Fvoronoi_correction} inferring
double-counting of surfaces and thus overestimating of the total cell
surface.

\begin{figure*}
\begin{tabular}{cc}
\begin{minipage}{0.47\linewidth}
\includegraphics[height=4cm]{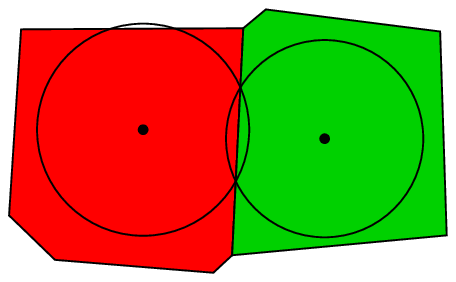}
\end{minipage}
&
\begin{minipage}{0.47\linewidth}
\includegraphics[height=4cm]{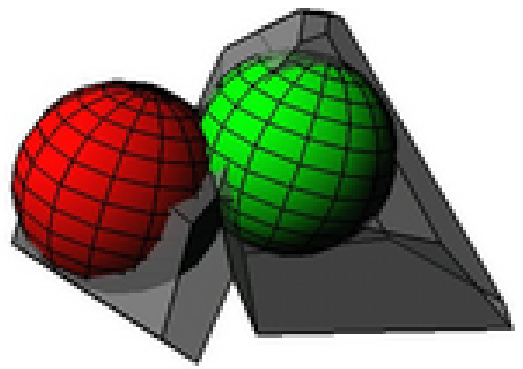}
\end{minipage}
\end{tabular}
\caption{\label{Fvoronoi}Visualization of two intersecting circles (spheres) and their corresponding Voronoi domains in
        two (three) dimensions. Position and orientation of the
        Voronoi contact line (plane) coincides with the circle
        (sphere) intersection. The Voronoi surfaces are also determined by the positions of other cells (not shown here).}
\end{figure*}

The weighted Voronoi tessellation \cite{schaller2004,okabe2000} of a set of spheres $\{\Vektor{r_i}, R_i\}$ 
\bea\label{Evoronoi}
V_i(t) = \{ \Vektor{x} \in {\mathrm R}^n \; &:& \; \left[\Vektor{x} -
\Vektor{r_i}(t)\right]^2 - R_i^2(t) \le \left[\Vektor{x} - \Vektor{r_j}(t)\right]^2 - R_j^2(t)\nn\\ 
        && \forall j \neq i\}
\eea
divides space into Voronoi regions -- convex polyhedra that may in some sense be associated with the space occupied by cell $i$ 
(see figures \ref{Fvoronoi_correction} and \ref{Fvoronoi}). 
This correspondence however is deceptive, as one can easily show that equation (\ref{Evoronoi}) leads to infinitely large intercellular contact surfaces 
at the boundary of the convex hull of the points  $\{\Vektor{r_i}\}$. 
In addition, in the case of a low cellular density the surfaces and volumes defined by the purely geometric approach (\ref{Evoronoi}) will evidently
overshoot the actual cellular contact surfaces and volumes by orders of magnitude.
On the other hand, Voronoi contact surfaces have been shown to approximate the cell shape in tissues remarkably well -- at least in two-dimensional
cross-sections \cite{honda1978}.
Therefore, in order to have a contact surface estimate valid for different modeling environments we use a combination of the two approaches by setting
\bea
A_{ij} = \min\left\{A_{ij}^{\rm sphere}\,,A_{ij}^{\rm Voronoi}\right\}\,.
\eea
In order to use the Voronoi surface, cells do not only
have to be in contact, but the Voronoi contact surface must be smaller
than the spherical contact surface, which can be the case for multiple cell
contacts, compare figure \ref{Fvoronoi_correction}. 
This combination leads to upper bounds of intercellular contact 
surfaces on tissue boundaries and preserves the Voronoi surfaces
within dense tissues by yielding a continuous transition between the
two estimates. The underestimation of the repulsive
forces in dense tissues within the Hertz model is in parts compensated
by using the Voronoi-based decreased
adhesive forces thereby leading to an increased net repulsion.
Depending on the local cellular deformations the difference
between the spherical and the Voronoi contact surface
can be in the range of 30\% within dense tissues.

Note that equations (\ref{Ehertz}) and (\ref{Ereclig1}) allow for different cell types by introducing varying radii, elastic moduli and receptor and ligand concentrations.
All forces act in the direction of the normals to the next neighbors and on the center of the spheres.
The total force on the cell $i$ is then determined by performing a sum over the next neighbors
${\Vektor{F}_i = \sum_{j \in {NN}(i)} \left(\Vektor{F}_{ij}^{\rm ad} -
\Vektor{F}_{ij}^{\rm el}\right) \cdot \Vektor{n}_{ij}}$
and in addition we record the sum of the normal tensions 
\bea
P_i = \sum_{j\in {NN}(i)} \frac{\abs{\Vektor{F}_{ij} \cdot 
        \Vektor{n}_{ij}}}{A_{ij}}\,,
\eea
where $\Vektor{n}_{ij}$ denotes the unit vector pointing from cell
$i$ to cell $j$.
The list of next neighbors is efficiently provided by 
the Delaunay triangulation.
Once a force has been calculated, the corresponding spatial step can
be computed from the equations of motion \cite{palsson2001,dallon2004}
\bea\label{Eeom}
m_i \ddot{r}_i^\alpha(t) = {F}_i^\alpha(t) 
        - \sum_\beta \gamma_i^{\alpha\beta} \dot{r}_i^\beta(t)  
        - \sum_\beta \sum_j \gamma_{ij}^{\alpha\beta} \left[\dot{r}_i^\beta(t) - \dot{r}_j^\beta(t)\right]\,,
\eea
where the upper Greek indices $\alpha, \beta \in \{0, 1, 2\}$ denote the
coordinates and the lower Latin indices $i,j \in \{0, 1, \ldots, N-1\}$ the
index of the cell under consideration. The adhesive or repulsive
forces as well as possible random forces on cell $i$ are contained in
the term $F_i^\alpha$, whereas the coefficients
$\gamma_i^{\alpha\beta}$ and $\gamma_{ij}^{\alpha\beta}$ represent
cell-medium and cell-cell friction, respectively.
A common isotropic choice for cell-medium friction is the normal Stokes relation
\bea
  \gamma_i^{\alpha\beta, \rm visc} = 6 \pi \eta R_i \delta^{\alpha\beta}\,,
\eea
which describes the friction of a sphere with radius $R_i$ within a
medium of viscosity $\eta$.

Most tissue simulations use the over-damped approximation 
$m_i {\ddot r}^\alpha_i(t) \approx 0 \qquad \forall i,\alpha, t$,
which is an adequate approximation for cell movement in medium
\cite{howard2001}, since the estimated Reynolds-numbers are
extremely small \cite{dallon2004}. 
Evidently, since additional adhesive bindings are at work, cellular
movement in a tissue is even more damped \cite{drasdo2003a}. In the
over-damped approximation, equation (\ref{Eeom}) reduces to a $3N
\times 3N$ linear system, that is sparsely populated and therefore can
in principle be solved using an iterative method \cite{dallon2004}. 
However, the large number of cells involved in larger multicellular tumor
spheroids would make this approach inefficient
-- both in terms of storage and execution time --
and limits the simulations to $\order{10^5}$ cells.
It is also not clear whether this intercellular drag
force term significantly contributes.
We have omitted this term and compensate for this by
a modified friction model which respects 
that the movement of bound cells is considerably inhibited.
In addition, one should keep in mind that within dense tissues many
intercellular contacts are mediated by the extracellular matrix (with
zero velocity). Such a friction term will rather contribute to the
diagonal part of the dampening matrix.
Therefore, we chose to approximate the term with the velocity
differences by increasing the isotropic cell-medium friction
coefficient by another term, i.~e., 
$\gamma_i^{\alpha\beta} = \gamma_i^{\alpha\beta, \rm
visc} + \gamma_i^{\alpha\beta, \rm ad} = \gamma_i \delta^{\alpha\beta}$ with
\bea\label{Ereclig2}
\gamma_i^{\alpha\beta, \rm ad} &=& \gamma^{\rm max} \delta^{\alpha\beta}
        \sum_{j \in { NN}(i)} A_{ij}
        \frac{1}{2}\left(1- \frac{\Vektor{F}_{i} \cdot \Vektor{n}_{ij}}{\abs{\Vektor{F}_{i}}}\right)\times\nn\\ 
        &&\times\frac{1}{2} \left(c_i^{\rm rec} c_j^{\rm lig} + c_i^{\rm lig} c_j^{\rm rec}\right)\,,
\eea
as illustrated in figure \ref{Fadhesion_friction}. 
Note that the above ansatz for the friction coefficient
scales with the intercellular contact surfaces and therefore
cells having many bounds to next neighbors will move less than unbound
cells. 
This is not an isotropic choice, since the forces
contribute to its calculation. With using these approximations, the system (\ref{Eeom})
becomes diagonal, i.~e.~ one has
\bea\label{Eeom2}
\Vektor{\dot r}_i = \frac{\Vektor{F}_i}{\gamma_i}\,.
\eea
As an option the model is capable
of including random
forces in order to mimic random cellular movement. 
However, the corresponding physiologic cellular diffusion
coefficients are in the range of $\order{10^{-4} \rm \mu m^2/s}$,
which leads to small displacements only. In the case of growing tumor
spheroids, the proliferation-driven tumor front will generally
overtake cells that have separated due to random movements. 
The stochastic nature contained in the
mitotic direction and the duration of the cell cycle obviously
suffices to yield isotropic tumor spheroids.
The simulations shown here have therefore been
performed without an additional stochastic force, unless otherwise noted.


\subsection{The cell cycle}\label{SScellcycle}

\begin{figure*}
\begin{tabular}{cc}
\begin{minipage}{0.47\linewidth}
\includegraphics[height=4cm]{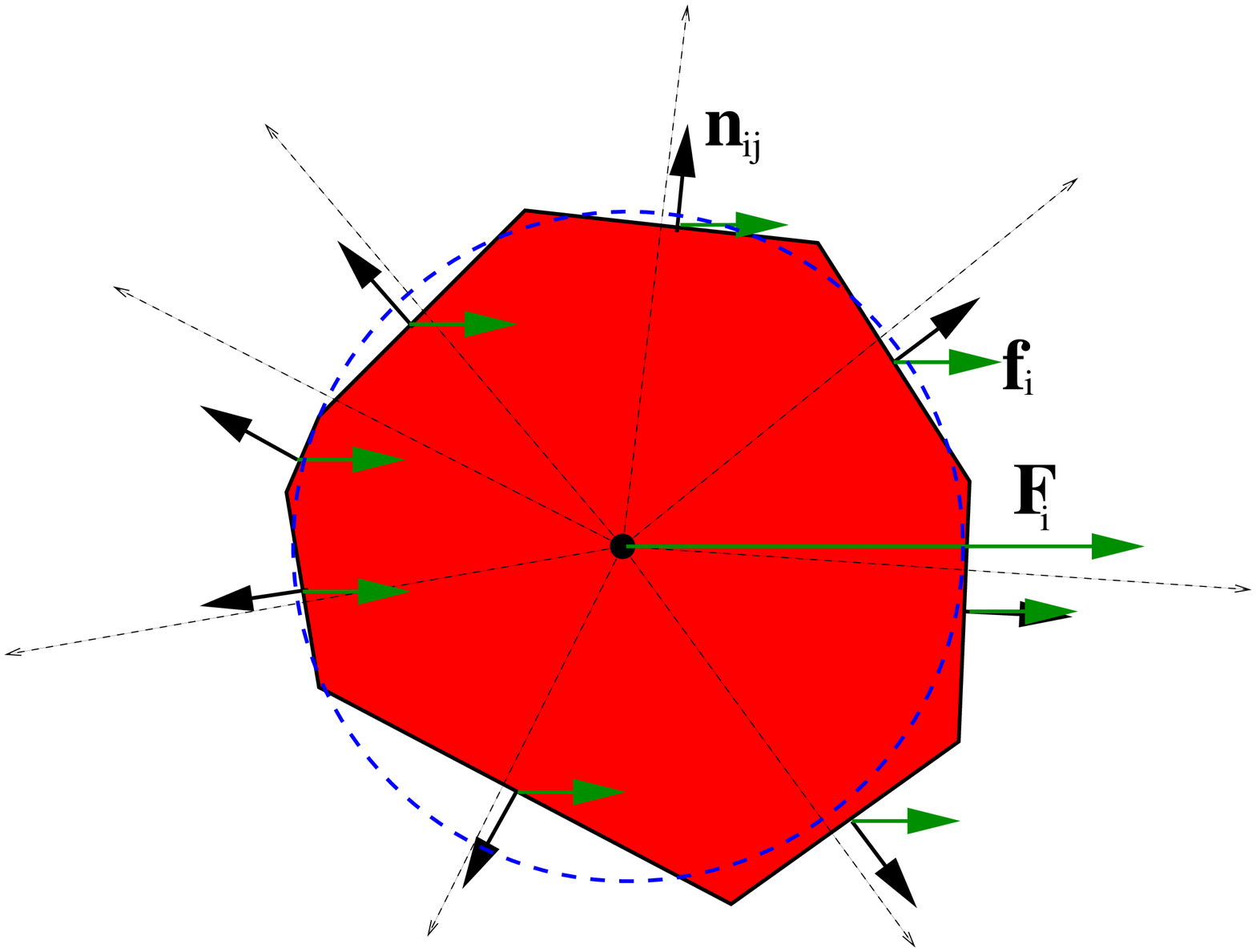}
\end{minipage}
&
\begin{minipage}{0.47\linewidth}
\includegraphics[height=4cm]{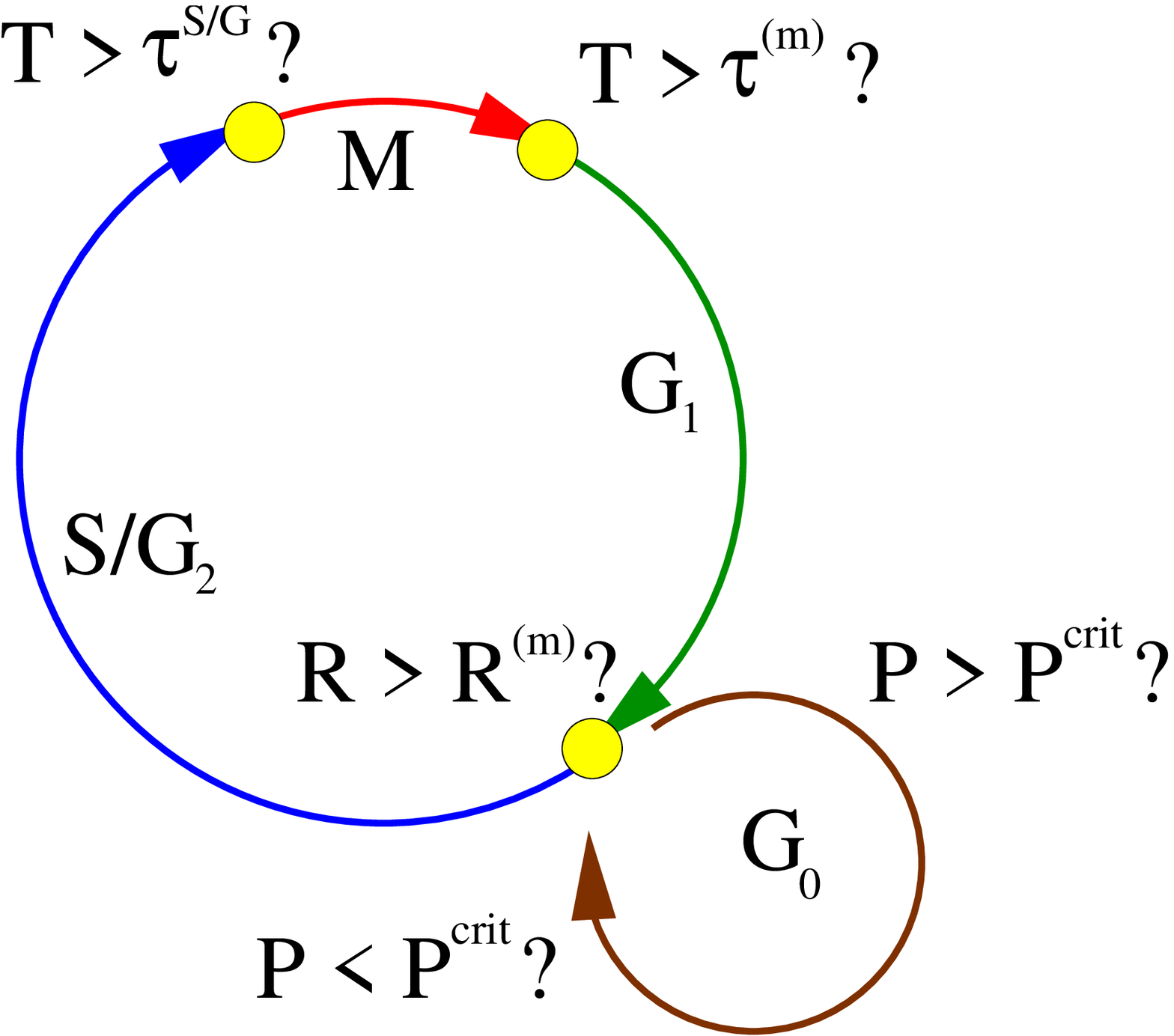}
\end{minipage}\\
\begin{minipage}{0.45\linewidth}
\caption{\label{Fadhesion_friction}The extent to which adhesive bonds contribute to friction depends on the direction of movement and on the contact surfaces.
        If total force and normal vector are parallel, the corresponding contact surface will not contribute at all to the friction coefficient in equation (\ref{Ereclig2}),
        whereas the contribution will be strongest with force and
        normal vector being antiparallel.}
\end{minipage}
&
\begin{minipage}{0.45\linewidth}
\caption{\label{Fcyclesketch}During cell division, cells reside in the
M-phase for $\tau^{(m)}$. Afterwards, the cell volume increases at a
constant rate in the $\rm G_1$-phase, until the pre-mitotic radius
$R^{\rm (m)}$ has been reached. At the end of the $\rm G_1$-phase, the
cell can either continue the cell cycle or enter the $\rm G_0$-phase, if
the normal tension $P_i$ exceeds a threshold. The ${\rm S}/{\rm G}_2$-phase lasts
for a time $\tau^{\rm S/G_2}$, after which mitosis is deterministically
initiated. The necrotic state can be entered at all times in the cell
cycle.}
\end{minipage}
\end{tabular}
\end{figure*}

In our model, cells have different internal states, which we chose to
closely follow the cell cycle in order to make 
comparisons with experimental data as intuitive as possible. Consequently, 
the cellular status determines the actions of the cellular agents.
We distinguish between 5 states: $\rm G_1$-phase, $\rm S/G_2$-phase,
M-phase, $\rm G_0$-phase, and necrotic, see also figure \ref{Fcyclesketch}.

During $\rm G_1$-phase, the cell volume grows at a constant rate
$r_{\rm V}$,
i.~e. the radius increases according to $\dot{R} = \left(4 \pi
R^2\right)^{-1} r_{\rm V}$, until the cell reaches its final mitotic radius
$R^{\rm (m)}$. The volume growth rate $r_{\rm V}$ is deduced by
assuming that cell growth is only performed during during $\rm G_1$-phase
\bea    
r_{\rm V} = \frac{2\pi \left(R^{\rm (m)}\right)^3}{3 \tau_{\rm G_1}}\,,
\eea
where $\tau_{\rm G_1}$ can be deduced from the minimum observed cycle
$\tau^{\rm min}$ time and the durations of the $\rm S/G_2$-phase and the M-phase.
Afterwards, no further cell growth is performed.
At the end of the $\rm G_1$-phase a checkpointing mechanism is performed 
where the cell can switch into $\rm G_0$-phase.
If the cellular tension exceeds the threshold $P^{\rm crit}$ at this position in the cell cycle, the cell enters the $\rm G_0$-phase, 
otherwise the cell enters the $\rm S/G_2$-phase.
Note that a different criterion for entering or leaving the $\rm
G_0$-phase would also be possible: Cells might enter $G_0$-phase at
any time in the cell cycle if
the local nutrient concentrations fall below thresholds or --
alternatively -- if toxic substances exceed certain
thresholds. 
In the present paper we
will restrict to interpreting cellular quiescence as contact
inhibition,
since there is experimental evidence that in case of EMT6/Ro cells
quiescence is not induced by lack of nutrients \cite{casciari1992,wehrle2000}.

During the S-phase the DNA for the new cell division is synthesized, whereas during $\rm G_2$-phase the quality of the 
produced DNA is controlled. In our model we do not distinguish between S-phase and $\rm G_2$-phase.
At the beginning of the phase the individual phase duration is determined using a normally-distributed random number generator \cite{gammel2003}
with a given mean and width. After this individual time has passed, the cells enter mitosis.

At the beginning of the mitotic phase -- which lasts for about half an hour for most cell types -- a mother cell divides and 
is replaced by two daughter cells. In the model these are slightly displaced in random direction, see subsection \ref{SSproliferation}. 
Afterwards the daughter cells are left to their initially dominating repulsive forces (\ref{Ehertz}).
As in the $\rm S/G_2$-phase the individual duration of the M-phase is determined using a normally-distributed random number generator.
Afterwards the daughter cells enter the $\rm G_1$-phase thus closing the cell cycle.
Note that we do not differentiate between the internal phases of mitosis.

During $\rm G_0$-phase, the cellular tension is monitored. Cells re-enter the cell cycle where they left it (i.~e.~at the beginning of the $\rm S/G_2$-phase)
if the cellular tension falls below the critical threshold $P^{\rm crit}$. Similar to the $\rm S/G_2$-phase no growth is performed.
Therefore in our model, the difference between the $\rm S/G_2$-phase and the $\rm G_0$-phase is that the duration of the first is
determined by the normally distributed individual time, whereas for
the duration of the latter the cellular tension is the determining factor.
Consequently, the cells in $\rm G_0$-phase can serve as a reservoir of cells ready to start proliferating as soon as there is enough space available, 
which is common to many wound-healing models \cite{galle2005a}.

Intuitively, cells enter necrosis as soon as the nutrient concentration at the cellular position falls below a critical threshold.
We study different mechanisms for the induction of necrosis within the
model and will be able to rule out possible candidates (see subsection
\ref{SSpopdyn}). Naturally, necrotic cells do not consume any nutrients and do slowly
decay. In our model this is represented by removing these cells from the
simulation at a rate $r^{\rm necr}$ -- without performing prior shrinking.

Note that the only stochastic elements involved so far are the direction of mitosis and the durations of the M-phase and $\rm S/G_2$-phase.
The first is required by the local assumption of isotropy, whereas the latter is required by the fact that proliferating cells having a common progenitor 
desynchronize rather quickly (usually after about 5 generations \cite{kreft1998}):
For these small systems of $\order{2^5}$ cells mechanisms such as nutrient
depletion or contact inhibition cannot explain the desynchronization.


\subsection{Proliferation}\label{SSproliferation}

\begin{figure*}
\begin{tabular}{cc}
\begin{minipage}{0.47\linewidth}
\includegraphics[height=4cm]{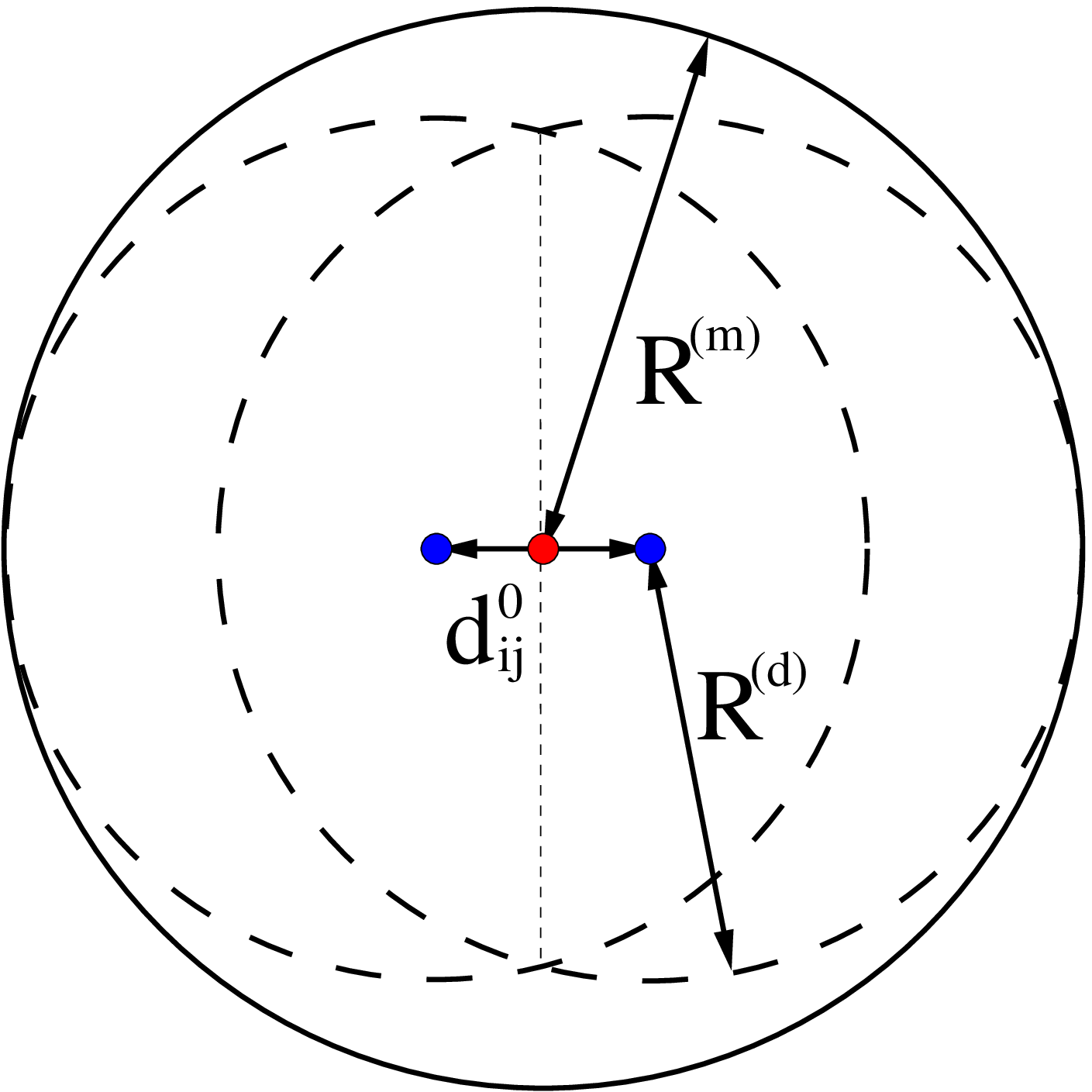}
\end{minipage}
&
\begin{minipage}{0.47\linewidth}
\includegraphics[height=4cm]{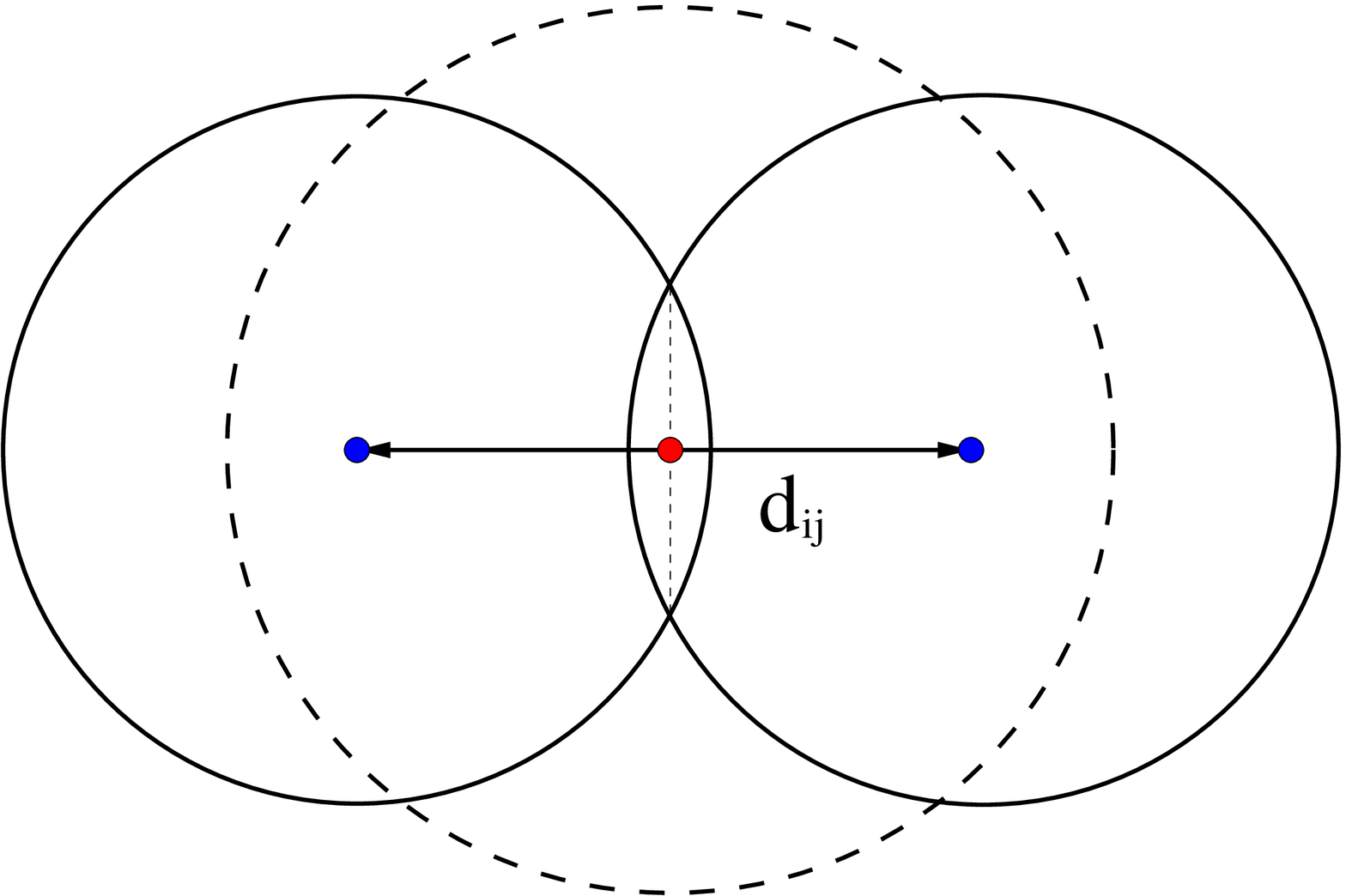}
\end{minipage}
\end{tabular}
\caption{\label{Fproliferation}
Illustration of the cell configuration right at proliferation
(left) and at the end of the M-phase (right).
At cell division, the radii of the daughter cells $R^{\rm (d)}$ 
are decreased to ensure conservation of the target volume during
M-phase. The resulting strong
repulsive forces drive the cells apart quickly. An adaptive timestep 
control ensures that the mitotic partners 
do not lose contact during M-phase. 
}
\end{figure*}

A cell will divide when the end of the $\rm S/G_2$ phase has been reached. 
The initial direction of mitosis is chosen randomly from a uniform
distribution on the unit sphere \cite{gammel2003}, which is the
simplest possible assumption.
Note however that since the cellular movement during the M-phase is not only determined by the mitotic partners but also
by the surrounding cells the effective direction of mitosis may generally change during M-phase -- depending on the configuration of the 
next neighbors.
The radii of the daughter cells are decreased $R^{\rm (d)} = R^{\rm
(m)} 2^{-1/3}$ to ensure conservation of the target volume during
M-phase and the daughter cells are placed at 
the distance $d_{ij}^0 = 2 R^{\rm (m)} (1-2^{-1/3})$ to ensure that
initially the deformations of surrounding cells do not change
drastically, see figure \ref{Fproliferation}.
One should be aware that at this stage the forces calculated in
equation (\ref{Ehertz}) cannot represent the actual mitotic separation
forces, since the considerable overlap $h = R^{\rm (m)} (2^{5/3} - 2)$
generates strong elastic forces in equation (\ref{Ehertz}) which has
then been applied far beyond its validity for small deformations. 
Therefore, to ensure for numerical stability, an adaptive step-size 
control has to be applied in the numerical solution of equation
(\ref{Eeom}) -- see the appendix -- since otherwise the contact
between the daughter cells might be lost immediately. Still, with an
adaptive timestep, the initial separation of mitosis will happen on a
timescale shorter than in reality.
To the sake of simplicity we will not use modified mitotic forces
within this article. One should keep in mind that the relative
shortness of the M-phase in comparison with
the complete cell cycle leads to a small fraction of cells being in the
M-phase. Therefore, we expect the consequences of our simplifying
assumption to be relatively small.

In figure \ref{Fproliferation} two cells are shown at proliferation and
right after the M-phase.
The bell-shape during mitosis resulting from the model is in
qualitative agreement with the physiologic appearance of mitosis.
One can also see that further
intercellular contacts may be lost, if the neighboring cells reside
perpendicularly to the direction of mitosis.
The direction of mitosis will generally
change during M-phase -- and thus considerably differ from figure
\ref{Fproliferation} right panel -- and thereby the temporarily lost
contact will in average be re-established, 
since the net forces will point to regions
of low cell density and thus lead to closure of gaps. At the boundary
of the spheroid however, cells may temporarily detach due to this
mechanism. Though this had not been intended, it does not seem in
contradiction to reality, since there exists experimental
evidence \cite{landry1981} that EMT6/Ro tumor spheroids loose cells at
the boundary due to mitotic loosening.
A macroscopic detachment of cells from the spheroid boundary has not
been observed in the simulation, since the spheroid growth velocity
has always been large enough to re-establish contact after some
time. However, such intermediate detachment events may very well
contribute to the overall apparent growth velocity.


\subsection{Nutrient consumption and Cell Death}\label{SSnutrients}

We view cells as bio reactors where oxygen and glucose react to waste products as lactose, water and carbon dioxide.
The clean combustion of glucose would require the molar nutrient uptake rate of oxygen to be 6 times the molar glucose uptake rate:
$C_6H_{12}0_6 + 6 0_2 \to 6 H_20 + 6 C0_2$.
However, for tumor tissue this cannot be the case as it is well-known that in the direct vicinity of 
tumors the concentration of lactic acid increases considerably which is a direct evidence for the incomplete combustion of glucose.
By experimental estimations of average oxygen and glucose uptake rates for another cell line a considerable deviation 
from the ideal ratio has been found with about 1:1
\cite{kunz_schughart2000}. For EMT6/Ro cells, in \cite{wehrle2000} a
ratio of about 1 : 3.9 is reported.

Thus, in our model all viable cells consume oxygen and glucose
diffusing in the surrounding extracellular matrix at specific but
constant rates.

The nutrient uptake rates can in principle depend on the cell type, the local concentration of both nutrients, the existence of internal cellular 
nutrient reservoirs and many other factors. 
However, few information about the qualitative dependence is known: 
most rates in the literature (see e.~g.~\cite{kunz_schughart2000}) 
are average values given in units of mol per seconds and volume of tissue since these data are 
obtained from whole cell populations without regard to the individual cell size, status and the local nutrient concentration.
In addition, the functional form of the dependence is unknown as well.
The simplest starting point is to assume that the nutrient uptake
rates only depend -- if at all -- on the local nutrient concentration.
For example, when dealing with a single
nutrient, quite often a Michaelis-Menten-like concentration-dependent
nutrient uptake rate is assumed, see e.~g. \cite{beuling2000}. This
however means the introduction of further parameters that may be
difficult to fix with the data available.

Depending on the cell type and on the local nutrient concentrations cells undergo apoptosis and/or necrosis when subject to 
nutrient depletion \cite{freyer1986a}.
In this specific application we choose necrosis as the dominant pathway to cell death and neglect the effects of apoptosis 
though there is experimental evidence that these processes are linked with each other \cite{bell2001}.
Necrotic cells are randomly removed from the simulation with a rate $r^{\rm necr}$. The effect of apoptosis in the simulation would be similar, though
apoptotic cells not break apart as necrotic cells but shrink and afterwards dissolve into small apoptotic bodies \cite{noble2003}. 
For the overall outcome of the total growth curve we expect insignificant changes by including apoptosis into the model.

With our computer simulation model we can test different hypotheses on
which critical parameters may influence the onset of necrosis:
For example, there could be two critical concentrations for both
oxygen and glucose or just one combined parameter with an unknown
dependence on the local concentrations.
In addition, there could also be other processes such as necrotic waste material inducing apoptosis and/or necrosis, which will not
be considered here.


\subsection{Nutrient distribution}\label{SSnutrientdist}

We consider the case of avascular tumor growth and therefore assume that the transport of nutrients is performed passively by diffusion.
Consequently, the diffusion through tumor tissue and also through the culture medium is described by a system of reaction-diffusion equations
\bea\label{Ediffusion}
\pdiff{u^{\rm ox/gluc}}{t} &=& \vec\nabla \left[ D^{\rm
ox/gluc}(\Vektor{x}; t)
\vec\nabla u^{\rm ox/gluc}(\Vektor{x},t)\right]\nn\\
        &&- r^{\rm ox/gluc}(\Vektor{x}; t)\,,
\eea
where $u^{\rm ox/gluc}(\Vektor{x},t)$ describes the local oxygen or
glucose concentration, $D^{\rm ox/gluc}(\Vektor{x}; t)$ the local
effective oxygen or glucose diffusion coefficient (which depends
implicitly on time via the cellular positions)
and $r^{\rm ox/gluc}(\Vektor{x}; t)$ the local oxygen or glucose
consumption rate. 
Though formally equation (\ref{Ediffusion}) might admit negative nutrient 
concentrations (even at low concentrations strong negative sink terms
may in principle exist), this can never happen in reality -- provided
the timestep is not too large: Cells will enter
necrosis (thereby stopping nutrient consumption) if the local
nutrient concentrations become too small. As the reaction rates depend
on the cellular status, they become implicitly dependent on the
nutrient concentrations, see also subsections \ref{SScellcycle} and
\ref{SSnutrients}.

In equation (\ref{Ediffusion}) we implicitly assume that the transport of matter can be described by an effective 
diffusion coefficient. 
This does not have to be the case, since cellular membranes pose complicated boundary conditions especially for larger
molecules such as glucose. In addition, convection may also contribute
to matter transport.
Only if the tissue is isotropic on scales larger than a cell diameter this assumption is justified.
Consequently, the discretization of equation (\ref{Ediffusion}) does
only make sense on lattices with spacings exceeding the cellular
diameters.

Though we use an effective diffusion coefficient $D_{\rm eff}$ it is sometimes necessary to allow for diffusivities varying on scales larger than
the cell diameter -- especially for larger molecules.
For example, the effective diffusion coefficient of glucose is about 700 $\rm \mu m^2/s$ in water, whereas it is only 100 $\rm \mu m^2/s$ in tissue \cite{casciari1988}.
This effect is less pronounced for smaller molecules such as oxygen with about 2400 $\rm \mu m^2/s$ in water and 1750 $\rm\mu m^2/s$ in tissue \cite{grote1977}.
Consequently, when modeling {\em in vitro} multicellular tumor spheroids one will have to take spatially varying diffusivities into account to appropriately model the
nutrient concentrations on the spheroid boundary. In our model, the
diffusion constant is set to measured tissue diffusivities in the
vicinity of cells and to the normal diffusivities in water anywhere
else. 
Therefore, by considering varying
diffusivities one is able to keep the rectangular shape of the
diffusion grid which is favorable for the numerical
solution, see also the appendix.
Note that a diffusion-depletion zone as in \cite{groebe1996} is
thereby automatically incorporated into the model. The difference is
that here the model does not a priori impose spherical symmetry. It can be checked
however by direct observation of the spherically-shaped nutrient
isosurfaces, that the rectangular shape of the boundary does not
greatly influence the nutrient distribution near the tumor.

Another possibility would be to solve the nutrient diffusion within
the spheroid only by assuming a spherical tumor symmetry with a time-dependent boundary moving with
the spheroid size. However, with such an approach the spherical symmetry would not be an outcome but an intrinsic ingredient of the model. 
Consequently, in such a model the spheroid shape would not be of any comparative value.

Equation (\ref{Ediffusion}) does only have a defined solution if the
initial conditions and the boundary conditions are set.
As in \cite{freyer1986a} it has been verified that the nutrient
concentration outside the tumor spheroid did not vary strongly between
the periodic refilling of nutrients, we approximated the experimental
system by imposing Dirichlet boundary conditions throughout the
simulation. The corresponding
initial and boundary concentrations have both been set to the
values used in the experiment.

\section{Results}\label{Sresults}


\subsection{Population Dynamics}\label{SSpopdyn}

The overall cell number is a parameter which can be quantified experimentally, either indirectly by simply calculating cell numbers
from observed tissue volumes or directly by extensive automated cell-counting.
In \cite{freyer1986a} the cell number has been determined indirectly for different concentrations of oxygen and glucose.
With our model we have calculated growth curves for different nutrient concentrations and different hypotheses of nutrient uptake and necrosis induction.
The simulations have been compared with four series of experimental data, where four different combinations of oxygen and glucose concentrations have been
investigated. Naturally, within one set of simulations all parameters
but the nutrient concentrations have been kept fixed.

\begin{figure*}
\begin{tabular}{cc}

\begin{minipage}{0.47\linewidth}
\includegraphics[height=7cm]{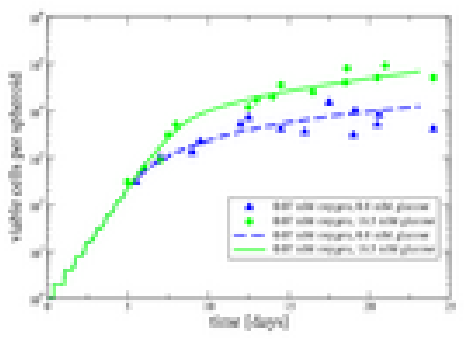}
\end{minipage}
&
\begin{minipage}{0.47\linewidth}
\includegraphics[height=7cm]{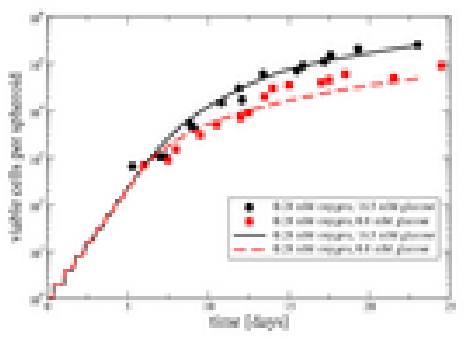}
\end{minipage}
\end{tabular}

\caption{\label{F_oxgluc_fit}
        Number of viable cells per spheroid for 0.8 and 16.5 mM
        glucose concentrations and either 0.07 mM oxygen concentration
        (left panel) or 0.28 mM oxygen concentration (right panel).
        Experimental data (symbols) were read off from \cite{freyer1986a}, whereas lines correspond to the computer simulations.}
\end{figure*}

We have tested the possibility that there exist critical
concentrations for the two nutrients separately. However, in this case 
either the glucose or oxygen concentration dominantly limit the cell
population dynamics. This does not reproduce the experimental data
\cite{freyer1986a}, since the growth curves for one of the nutrient
concentrations being kept constant depend strongly on the
concentration of the other nutrient. Therefore, since low oxygen and large glucose
concentrations can result in similar population dynamics as large
oxygen and low glucose concentrations (\cite{freyer1986a}), both
concentrations must enter the critical parameter. 
We have also tested
the possibility of concentration-dependent nutrient consumption rates
with the functional form of the Michaelis-Menten type kinetics 
\bea
r^{\rm nut} = r^{\rm min} + \frac{(r^{\rm max} - r^{\rm min}) C^{\rm nut}}{C_{1/2} + C^{\rm nut}}\,.
\eea
This model however uses additional parameters that cannot be fixed with
the present data -- even when omitting $r^{\rm min}$.
In addition, the values for $C_{1/2}$ in the literature for oxygen-dependent
proliferation \cite{roose2003} of $0.0083$ mM point into the direction
that the oxygen consumption rates are always within the range
of saturation, since the local oxygen concentration has always been larger than
0.04 mM throughout the spheroids.
Consequently, we have assumed constant cellular
oxygen and glucose uptake rates for non-necrotic cells in the present model. We chose the product of 
oxygen and glucose concentration to be the limiting factor to sustain cellular viability.
This simple ansatz did suffice to reproduce the experimental cellular
growth curves (see figure \ref{F_oxgluc_fit}). The best
fit is achieved with the parameter set shown in table
\ref{Tparameters}. The corresponding tumor morphology is addressed in subsection
\ref{SSmorphology}.

Unfortunately, no error bars are given in \cite{freyer1986a} and the
experimental data scatter considerably even on a logarithmic
scale, see figure \ref{F_oxgluc_fit}.
Apart from the difficulty of establishing a defined experimental
system in biology, this large scatter is also due to the necessity of
destroying the spheroids during the measurements. Therefore, a whole
ensemble of spheroids had to be measured.
Since the monoclonality of these spheroids is not ensured, it is not a
priori clear whether a single spheroid might contain several species or whether 
different spheroids might belong to different species with individual
growth characteristics.
In order to employ a procedure to minimize deviations between the
simulation and experimental data we defined estimated error bars by 
calculating the difference to the artificial Gompertz growth curve
\bea\label{Egompertz}
N(t) = N_0 \exp\left[ \frac{\alpha}{\beta}\left(1 - e^{-\beta t} \right) \right]\,,
\eea
which is known to reproduce most growth processes in nature with
remarkable accuracy \cite{ling1993}.

Not every hypothesis on nutrient consumption and necrosis induction
leads to acceptable agreement with experimental data -- indicating the
sensitivity of the model.
The theoretical predictions lie within the scattering region, see figure \ref{F_oxgluc_fit}.
Qualitatively, one can see that for all the simulations the initial exponential growth phase soon enters a crossover to a polynomial growth.
In our model this crossover is due to two distinct mechanisms -- contact inhibition and nutrient depletion --  which lead to the similar outcome that after a certain
time dominantly the spheroid surface will contribute to the
proliferation, i.~e.
\bea
\tdiff{N}{t} = \alpha N^{2/3}\,,
\eea
which has the polynomial solution $N(t) = N_0 \left[ 1 + \beta t + \beta^2 t^2/3 + \beta^3 t^3/27 \right]$ with $\beta=\alpha/N_0^{1/3}$ \cite{mueller_klieser2002}.
Apart from the fact that necrosis is evidently more likely when 
nutrients are rare, the mechanisms cannot be clearly 
distinguished with a glance at the total growth curves in figure
\ref{F_oxgluc_fit}. 
Even in the case where both nutrients are rare,
the growth curve can be fitted by above equation: The scatter of the
data does not allow to exclude this possibility. However, given that
tumor spheroids saturate at a certain size, the above model cannot be
valid in all regimes of tumor growth.

\begin{figure*}
\begin{tabular}{cc}
\begin{minipage}{0.47\linewidth}
\includegraphics[height=7cm]{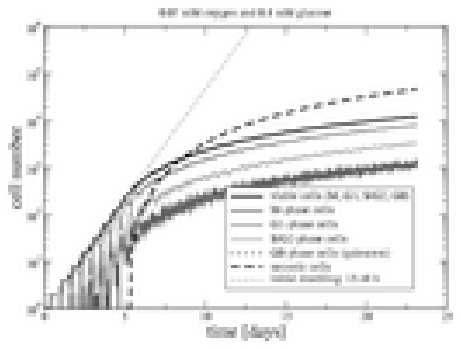}
\end{minipage}
&
\begin{minipage}{0.47\linewidth}
\includegraphics[height=7cm]{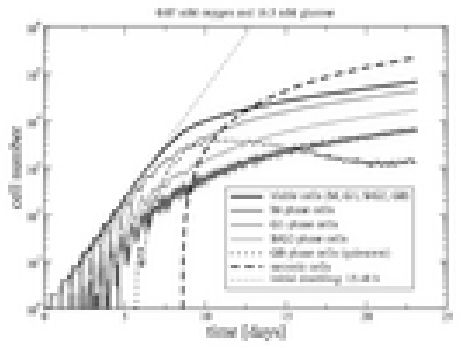}
\end{minipage}\\

\begin{minipage}{0.47\linewidth}
\includegraphics[height=7cm]{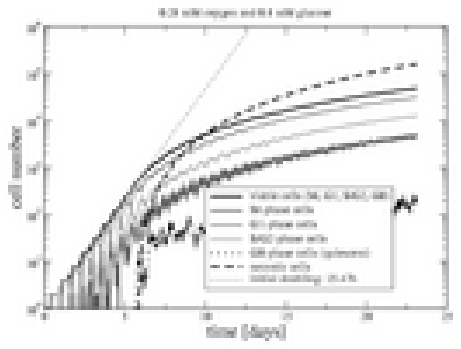}
\end{minipage}
&
\begin{minipage}{0.47\linewidth}
\includegraphics[height=7cm]{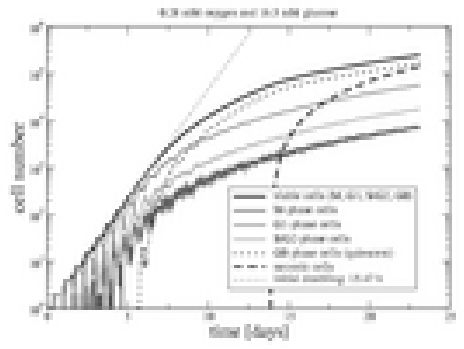}
\end{minipage}
\end{tabular}
\caption{\label{Fccycle}
        Cell cycle distribution for different oxygen and glucose concentrations.
        Depending on the external nutrient concentrations, significant 
        differences mark the dominance
        of different mechanisms to limit the cell cycle. Fits to the 
        regions of exponential growth -- marked by the complete
        absence of necrotic and quiescent cells -- reproduce the
        shortest observed cycle time within statistical
        fluctuations. 
        The initial oscillations in the sub-populations in the cell
        cycle stem from the fact that the cells divide synchronously
        at the beginning -- their frequency is the inverse cell cycle
        time. 
        After each cell division, the daughter cells
        draw new duration times for the $\rm S/G_2$-phase and the
        $M$-phase from a Gaussian distribution, compare table
        \ref{Tparameters}, which leads to a dampening of the
        oscillations and finally to complete desynchronization of cell division.
        The occurrence of contact inhibition or necrosis increases the dampening effect,
        since the migration through the cell cycle is impaired.
        Note that in the case of few nutrients
        contact inhibition does not play a role, as there are no
        quiescent cells (top left).}
\end{figure*}

Since the mechanism of contact inhibition leads to cells 
resting in $\rm G_0$ rather than cells 
entering necrosis the differences can easily be analyzed in the cell 
cycle distribution.
In figure \ref{Fccycle} it is evident that for 0.07 mM oxygen and 0.8
mM glucose concentrations (upper left panel) the nutrient starvation
is the dominant limiting factor to cell cycle inhibition, since there 
are nearly no cells in $\rm G_0$-phase and the majority
of cells is necrotic. In the case of nutrient abundance 
(0.28 mM oxygen and 16.5 mM glucose, figure \ref{Fccycle} lower right panel) 
however, the majority of cells resides in $\rm G_0$-phase during days
6-23, which is an indication for contact inhibition being the dominant
reason for the crossover, as is also assumed in
other models \cite{galle2005a,drasdo2003a}.
This is also confirmed by the cross-sections of the computer simulated tumor spheroids, see figure \ref{Fcross_section}.
Though in the case of nutrient abundance necrosis sets in much later, the number of necrotic cells rises
at a much stronger slope and it is to be expected that necrosis will displace the contact inhibition as the major cause for 
surface-dominated growth after 25 days (with overall roughly $5 \times 10^5$ cells involved, the simulations become very extensive and memory-consuming).
Such a displacement of dominating mechanisms is already visible for some intermediate nutrient concentrations. For example, in the case of
$0.07$ mM oxygen and $16.5$ mM glucose concentrations the number of cells in $\rm G_0$-phase first rises to reach its maximum after 10 days and 
afterwards decays in combination with a strong rise in necrotic cells
(figure \ref{Fccycle}, upper right panel).
Such a behavior is not observed in the regime of large oxygen and low
glucose concentrations (figure \ref{Fccycle} lower left panel), where
necrosis and contact inhibition set in
simultaneously and nutrient starvation is the main limiting factor. 
This is due to the considerably decreased glucose diffusion coefficient in tumor tissue, whereas the 
diffusion coefficient of oxygen is nearly the same in tissue and
water, compare subsection \ref{SSnutrientdist}.
Consequently, the already low glucose concentration 
of 0.8 mM at the boundary drops rapidly when the
number of tumor cells increases, since new glucose supply diffuses very slow from the outside.


\subsection{Tumor Spheroid Morphology}\label{SSmorphology}

\begin{figure*}
\begin{tabular}{cccc}
\begin{minipage}{0.2\linewidth}
\includegraphics[width=4.0cm]{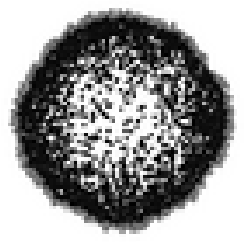}
\end{minipage}
&
\begin{minipage}{0.2\linewidth}
\includegraphics[width=4.0cm]{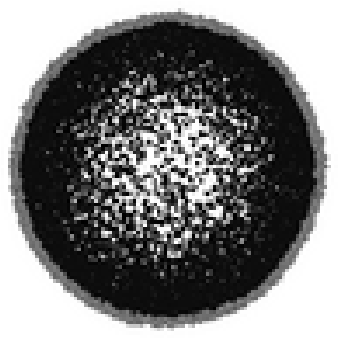}
\end{minipage}
&
\begin{minipage}{0.2\linewidth}
\includegraphics[width=4.0cm]{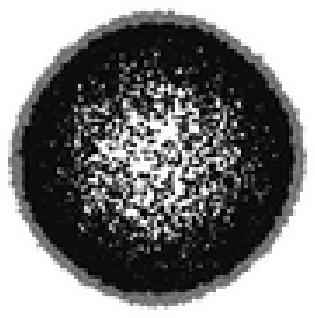}
\end{minipage}
&
\begin{minipage}{0.2\linewidth}
\includegraphics[width=4.0cm]{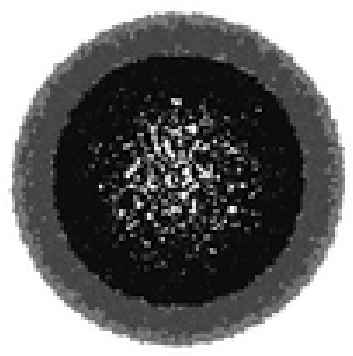}
\end{minipage}
\\
\begin{minipage}{0.2\linewidth}
\includegraphics[width=4.0cm]{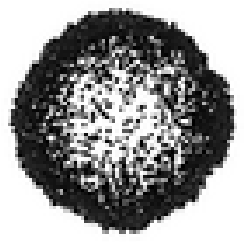}
\end{minipage}
&
\begin{minipage}{0.2\linewidth}
\includegraphics[width=4.0cm]{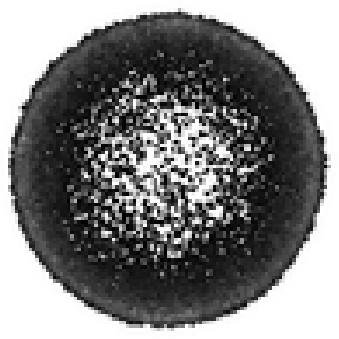}
\end{minipage}
&
\begin{minipage}{0.2\linewidth}
\includegraphics[width=4.0cm]{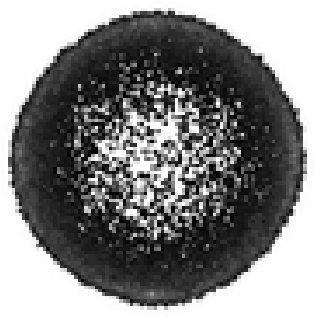}
\end{minipage}
&
\begin{minipage}{0.2\linewidth}
\includegraphics[width=4.0cm]{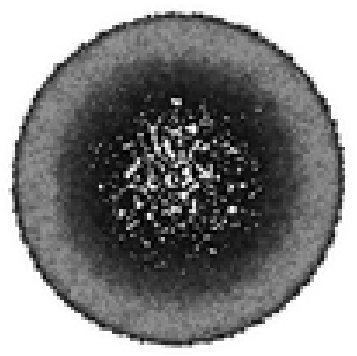}
\end{minipage}
\end{tabular}
\caption{\label{Fcross_section}Cross-section of computer-simulated tumor spheroids after $23$ days of simulation time. The first row shows the cellular status 
        (necrotic cells painted black, quiescent cells in dark grey, cells
        in the cell cycle in lighter shades of grey), whereas the second row visualizes the cellular tension
        (free cells painted black, cells under strong pressure in light
        grey). Nutrient concentrations from left to right are given by 
        $0.07$ mM oxygen and $0.8$ mM glucose, $0.07$ mM oxygen and $16.5$ mM glucose,
        $0.28$ mM oxygen and $0.8$ mM glucose, and $0.28$ mM oxygen and $16.5$ mM glucose.}
\end{figure*}

To estimate the quality of a mathematical model one has to find experimentally accessible parameters.
This is especially difficult when thinking about tissue morphology, since very often the patterns are hard to quantify in terms of numbers.
The morphology of three-dimensional tumor spheroids is rather simple: An inner necrotic core is surrounded by a layer of quiescent cells,
which is in turn surrounded by the outer layer of proliferating cells.
Qualitatively, this morphology is well reproduced in the case of
initial nutrient abundance, see upper right panel in figure
\ref{Fcross_section}.
In the case of nutrient starvation however there is virtually no layer
of quiescent cells (figure \ref{Fcross_section} upper left panel), 
as contact inhibition is not of importance in this scenario (see
figure \ref{Fccycle} 
upper left panel). 
This would be different if quiescence is 
induced by nutrient limitations: In this case, the necrotic core would
always be surrounded by a layer of quiescent cells.
Indeed, experimental observations \cite{casciari1992} suggest
that neither nutrient depletion nor the related acidic pH induce the
cellular quiescence.
It is evident from figure \ref{Fcross_section} that the size of the layers depends on the boundary concentrations.
In addition, it also depends on the nutrient consumption rates and diffusivities of oxygen and glucose within the tumor tissue.
The size of the necrotic core is also very sensitive on the rate at
which necrotic cells are being removed from the simulation.

Note that in the spheroid cross-sections it is evident that -- if oxygen and/or glucose are limited -- 
a relatively small number of cells with constant nutrient uptake rates suffices to drop the nutrient levels under the critical threshold thus
leading to the onset of necrosis and the absence of a layer of quiescent cells in the end of the simulations, compare also figure \ref{Fccycle}.
This is different for a model with concentration or cell-cycle dependent nutrient uptake rates.
In the first case the absolute value of the nutrient concentration gradients would be decreased thus giving rise to a broader viable layer which
-- in turn -- could allow for the existence of a quiescent layer.
In the second case the intermediate emergence of cellular quiescence (see figure \ref{Fccycle}) would also decrease the absolute value of the
nutrient concentration gradient towards the necrotic core, which would prolong and eventually stabilize the existence of a quiescent layer 
also for nutrient depleted configurations.
Therefore, in order to distinguish between nutrient uptake models, the tumor spheroid morphology is an important criterion, whereas the simple
total growth curve is not sufficient to make quantitative predictions
about the mechanisms at work.

Interestingly, the spheroids in figure \ref{Fcross_section} are fairly round, especially for the case where nutrients are provided in abundance.
This is due to the stochastic nature of the mitotic direction which forces initial differences to average out after some time -- which can easily be
verified by restarting the computer code with similar parameters but different seed values for the random number generator (data not shown).
This is in agreement with many spheroids observed in the experiment
\cite{freyer1986a} and in other computer simulations
\cite{drasdo2003a}.
However, the spheroids are less spherical for extreme nutrient
depletion, since firstly the small cell number yields less
stochastic events that contribute to the averaging and secondly the
emergence of localized holes in the necrotic core is not
counterbalanced by a strong mainly isotropic proliferative pressure
from the proliferating rim -- as is the case for large nutrient
concentrations.
The sometimes observed deviations from the spherical form
\cite{freyer1986a} can also have additional reasons:
The spheroids might be hetero-clonal while all cells in our simulation
are assumed to be monoclonal.
If a spheroid does not develop from a single but two genetically differing cells, these cells might exhibit different growth characteristics.


\subsection{Parameter Dependence}\label{SSparamdep}

The growth curves shown in figure \ref{F_oxgluc_fit}
have been calculated using the -- comparably many -- parameters in table \ref{Tparameters}.
\begin{table*}
\begin{minipage}{0.9\linewidth}
\begin{tabular}{c|cc|c}
parameter & value & unit & comment\\
\hline
ECM viscosity $\eta^{\rm VISC}$ & $5 \cdot 10^{-3}$ & $\rm kg/(\mu m s)$ &
\cite{galle2005a}, estimate\\
adhesive friction $\gamma^{\rm max}$ & 0.1 &    $\rm kg/(\mu m^2 s)$ &
\cite{galle2005a}, estimate\\
receptor concentration $c^{\rm rec}$ & 1.0 & \# & fixed\\
ligand concentration $c^{\rm lig}$ & 1.0 & \# & fixed\\
oxygen diffusivity $D_{\rm eff, ox}^{\rm tissue}$ & 1750.0 & $\rm\mu m^2/s$ & \cite{grote1977}\\
glucose diffusivity $D_{\rm eff, gluc}^{\rm tissue}$ &  105.0 & $\rm\mu m^2/s$ & \cite{casciari1988}\\
mitotic phase $\tau^{\rm (m)}$ & $(3.6 \pm 0.9) \cdot 10^3$ & s & estimate\\
$\rm S/G_2$-phase $\tau^{\rm S/G_2}$ & $(18.0 \pm 7.2) \cdot 10^3$ & s & estimate\\
shortest cycle time $\tau^{\rm min}$ & $54.0 \cdot 10^3$ & s &
\cite{landry1981,freyer1986a,casciari1992}, 
estimate\\
mitotic cell radius $R^{\rm (m)}$ & $5.0$ & $\rm \mu m$ & estimate\\
cell elastic modulus $E$ & $1.0 \cdot 10^{-3}$ & MPa & \cite{galle2005a}, estimate\\
cell Poisson number $\nu$ & $0.5$ & \# & assumption\\
adhesive coefficient $f^{\rm ad}$ & $1.0 \cdot 10^{-4}$ & $\rm \mu N/\mu m^2$ & eq. overlap\\
necrosis absorption rates $r^{\rm necr}$ & $2.0 \cdot 10^{-6}$ & cells/s & estimate/fit\\
critical cell tension $P^{\rm crit}$ & $0.6 \cdot 10^{-3}$ & MPa & fit parameter\\
oxygen uptake $r^{\rm ox}$ & $20.0 \cdot 10^{-18}$ &  mol/(cell s) & fit parameter\\
glucose uptake $r^{\rm gluc}$ & $95.0 \cdot 10^{-18}$ & mol/(cell s) & fit parameter\\
critical product $p^{\rm oxgluc}$ & $0.025$ & $\rm mM^2$ & fit parameter\\
\end{tabular}
\caption{\label{Tparameters}Best fit model parameters that are used in
        the simulations shown in figures \ref{F_oxgluc_fit},
        \ref{Fccycle}, and \ref{Fcross_section}. See text
        for explanations.}
\end{minipage}
\end{table*}
However, since mainly deterministic and rather physically-motivated interactions are assumed, more parameters 
than in PDE or cellular automaton models can be accessed by
independent experiments and do not need to be varied as fit
parameters.
Some of these parameters deserve special attention:
The elastic parameters of EMT6/Ro tumor cells might differ from those
in our simulation, where incompressibility has been assumed -- see
table \ref{Tparameters}.
Assuming reduced Poisson ratios $\nu \approx 0.3$ and elasticities of
$E \approx 750$ Pa \cite{galle2005a,roose2003}, one may obtain deviations
in the elastic forces in (\ref{Ehertz}) in the range of up to 50
percent.
However, even with these different elastic constants the growth
characteristics does not change significantly: This is due to the fact
that the cellular tensions relax on a much shorter timescale than the
cell cycle time.
An initial cycle time of 17 h has been obtained in \cite{freyer1986a} using
a Gompertz fit to the spheroid volume. This fit had been applied to
already existing small spheroids that may exhibit growth retardation
effects. 
For cells that had separated at the spheroid boundary, a
cell cycle time of only 13 h \cite{landry1981} has been observed. 
Therefore -- and in order to reproduce the slopes correctly -- we have
used a slightly decreased shortest possible cycle time.
The cell tension defined here is simply a sum
over all normal tensions with the next neighbors. 
The value that we have obtained as fit parameter is about 6 times
as large as the critical cellular compression used as a criterion for
contact inhibition in similar simulations (90 Pa in \cite{galle2005a}).
In part, this may be due to the Voronoi surface correction -- surfaces
tend to be smaller than sphere surfaces -- which leads to generally
larger normal tensions.
The remaining discrepancy should be attributed to 
the fact that we use a different cell line and the inherent model
differences. 
The removal rate $r$ of necrotic cells did not have a considerable impact on
the macroscopic number of viable cells and the spheroid size.
However, it can also be seen in figure
\ref{Fcross_section} that due to the removal of necrotic cells holes
emerge. Then the mechanical coupling from the necrotic tumor core
towards the boundary will be disrupted. Therefore, for the used
elastic and adhesive parameters, the parameter $r$ 
mainly controls the number of necrotic holes in the center. 
Note that this is different however, in a scenario with considerably increased
adhesion, where the mechanical coupling is not disrupted and the rate
constant $r$ does have an influence on the spheroid size and thereby
on the overall cell number. 

In accordance with the assumption of contact inhibition being the 
dominant cause for the crossover from exponential to polynomial growth
in the case of nutrient abundance, the initial phases of the
theoretical growth curve for $0.28$ mM oxygen and $16.5$ mM are 
dominantly dependent on the critical cell tension, whereas the other 
growth curves -- especially for nutrient depletion -- strongly
depend on the nutrient uptake rates and the necrotic parameter.
Generally, the late stages of spheroid growth depend critically on the
nutrient-related parameters. The resulting parameters for nutrient
uptake rates are well within the range observed in the literature
\cite{casciari1992,wehrle2000,groebe1991,freyer1985,freyer1986a}, 
though some considerable variances even within the literature exist. 
Apart from the fact that mostly different cell lines are analyzed, the
additional problem exists that the values in the literature are
usually volume-related uptake rates that have been fitted on
experimental data. Consequently, the extracted cellular uptake rates
depend on the corresponding cellular packing density of these
systems.
It must be kept in mind that these rates represent average values over
the whole ensemble of cells present in the spheroid. For example,
quiescent cells could have a considerably-decreased nutrient uptake
rate. 
In addition, there is evidence that glucose uptake rates can be
related to the local concentration of available oxygen
\cite{casciari1992}.
The present quality of the data however does not allow to
discriminate between more sophisticated models 

Note that in the over-damped approximation of equation
(\ref{Eeom}) the solution is calculated as a ratio of 
combined elastic and adhesive forces to a friction parameter, which
is largely influenced by cell-cell adhesion.
Therefore, the model will not be very sensitive on the specific
adhesion coupling constants and the adhesion-determined friction, 
as rather their ratio is mainly influencing the model behavior as long
as elastic forces are small.


\subsection{Saturation of growth curves}\label{SSsaturation}

\begin{figure*}
\begin{tabular}{cc}
\begin{minipage}{0.47\linewidth}
\includegraphics[height=7cm]{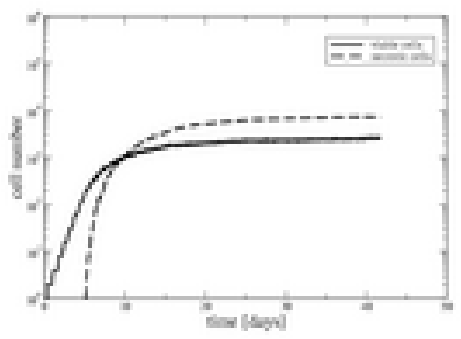}
\end{minipage}
&
\begin{minipage}{0.47\linewidth}
\includegraphics[height=7cm]{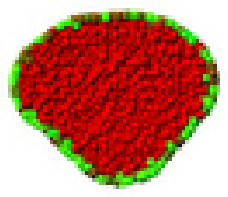}
\end{minipage}
\end{tabular}
\caption{\label{Fsaturation}
After assuming an increased adhesive coupling the emergence of  
holes within the necrotic core is completely inhibited.
In addition, the cells have been displaced randomly by a Gaussian
distribution with width $\Delta x_i = \sqrt{2 D \Delta t}$ at every timestep.
A steady-state flow equilibrium is established 
leading to approximate growth saturation of the spheroid (left) in the
observed time range. 
No further mechanisms need to be assumed.
In addition, the fast closure of holes in the necrotic core can lead
to deviations from the spherical symmetry (right). Cells in
the cell cycle are marked in light grey, quiescent cells in grey and 
necrotic cells are depicted by dark grey.
}
\end{figure*}

A complete saturation of the cell number or spheroid size -- as 
suspected by \cite{freyer1986a} and others \cite{folkman1973} -- cannot
be reproduced in the computer simulations with the parameters in table
\ref{Tparameters}.
The large scatter of the data in the case of
nutrient depletion (figure \ref{F_oxgluc_fit} left panel) does not exhibit a clear
saturation within 25 days, which is not reached in the other configurations anyway. 
For the explanation of a growth saturation the nature of the additional mechanism remains controversial. 
For example, in \cite{folkman1973} an effective movement of cells towards the necrotic core
has been observed leading to the assumption of a chemotactic signal secreted by necrotic cells.
The corresponding computer simulations in \cite{dormann2002} did lead to saturation.
Since it is somewhat arbitrary to assume that tumor cells follow a
necrotic signal we also tested a simpler hypothesis:

In figure \ref{Fcross_section} macroscopic holes are visible within the necrotic core -- created by the 
removal of necrotic cells from the simulation. Once such a hole is
established, it even tends to grow, since the intercellular adhesion 
is of very short range. (Recall that equation
(\ref{Ereclig1}) depends on the contact surface.) 
We have found that an increase of adhesive normal forces to 
$f^{\rm ad} = $ 0.0003 $\rm \mu N/\mu m^2$ suffices to close the
visible holes completely -- thereby inevitably coupling the
proliferating ring to
the necrotic core which finally leads to apparent growth saturation, compare
figure \ref{Fsaturation}. 
Note however, that in the presence of stochastic forces, complete
saturation (lasting infinitely long) can never be observed, since
already the seldom case of cells leaving the spheroid will lead to
further colonies that might recombine.
Consequently, the volume loss generated by removing
necrotic cells with a certain rate must be balanced by a movement of 
proliferating or quiescent cells from the outer layers into the
necrotic core. In addition, the outward component of the proliferative pressure on
the outer layer is counterbalanced by the increased cellular
adhesion as well. For such a system, a growth saturation is inevitable:
As in the late stages of spheroid growth the cellular birth rate
can be assumed to be proportional to the spheroid surface 
$R_{\rm birth} \approx \alpha N^{2/3}$ and the rate of 
cell removal is proportional to the number of necrotic cells residing 
in the center, the total cell number can be described by
\bea
\tdiff{N}{t} = \alpha N^{2/3}(t) - \beta\left[N(t) - \gamma N^{2/3}(t)\right]
\eea
with $\alpha, \beta, \gamma$ being positive constants.
Above equation resembles the growth law of Bertalanffy \cite{ling1993}.
The solution of this equation reaches the steady state
$N_\infty = \left(\frac{\alpha}{\beta} + \gamma\right)^3$, which is
stable for $\beta > 0$.
Therefore, in this regime the nutrient depletion is the dominant
factor limiting tumor spheroid growth.

We conclude that growth saturation of both cell number and spheroid
radius in off-lattice computer
simulations can be reached by assuming increased intercellular adhesion
forces.
In that case viable cells move towards the necrotic core (data not shown).
The assumption of some diffusing signal as in \cite{dormann2002} is
not necessary.
Interestingly, during the period of saturation, deviations from the
spherical shape can emerge: 
The position of unstable intermediate holes within the necrotic core is
randomly distributed and gives rise to macroscopic deviations from
spherical shape on the spheroid surface. Therefore, an irregular
spheroid shape can also be explained by individual durations
of the necrotic process.
Note that another candidate for a cell loss mechanism is shedding of cells at
the spheroid surface \cite{landry1981,landry1982}. 
All these mechanisms might could be combined with an
involvement of metabolic waste products in the induction of necrosis.


\section{Summary}\label{Ssummary}

We have demonstrated that the Voronoi/Delaunay hybrid model can very well be used to establish agent-based cell-tissue simulations.
The Voronoi/Delaunay approach provides some advantages:
Firstly, compared to the description of cells by deformable spheres, 
the Voronoi tessellation provides an improved estimate of
contact surfaces within dense tissues. 
The present model combines the advantages of both model concepts.
Secondly, the weighted Delaunay triangulation is an efficient method to
determine neighborship topologies for differently-sized sphere-like objects. 
In addition, it can efficiently be updated in the case of moving objects.
The model is very rich in features and therefore allows many comparisons with the experiments.
It can easily be combined with established models on cellular adhesion and elasticity that rely on direct experimental observables.
Therefore it allows some of its parameters to be fixed by independent experiments. 
The parameters which had to be determined with respect to macroscopic quantities represent existing physical quantities.
Since such quantities can be falsified in future experiments, the model provides predictive power to a greater extent than
differential equation or cellular automaton approaches.

Unlike previous models
\cite{dormann2002,drasdo2003a,groebe1996} -- which only
considered the influence of one nutrient on the dynamics of
three-dimensional multicellular tumor spheroids -- 
we were able to reproduce the experimental growth curves 
with a single parameter set by considering the spatiotemporal 
dynamics of both the oxygen and glucose concentrations simultaneously.
A saturation of growth could be obtained by increasing intercellular
adhesive forces threefold.

On the one hand, the typical spheroid morphology is reproduced
qualitatively very well.
On the other hand, a quantitative reproduction not only of cell
population growth curves but also of spheroid morphology 
could allow for a more detailed analysis of nutrient consumption
models:
For a different cell line an oxygen : glucose uptake ratio of about
1:1 has been found \cite{kunz_schughart2000}.
In contrast, our computer simulations point to the scenario that the oxygen 
consumption rates are much smaller (about 1:5) than the glucose
consumption rates (table \ref{Tparameters}), though the values are
within the ranges of uptake rates in the literature if considered
separately. 
This discrepancy may be due to several reasons.
Firstly, there is strong experimental evidence that 
the ratio of oxygen and glucose uptake in the case of EMT6/Ro cells
considerably differs even from the ratio of 1 : 1. For example, in
\cite{wehrle2000} a ratio of 1 : 3.9 is suggested.
Secondly, the effective diffusivities within tissue for oxygen and 
glucose obtained from \cite{grote1977} and \cite{casciari1988} might
not be correct -- this would lead to different currents of oxygen 
and glucose within the spheroid.
Thirdly, the model assumptions of roughly constant nutrient uptake
rates and the product of both concentrations being the critical
parameter for necrosis might not be correct.

We have seen that the quantitative analysis of the overall growth curve
can in principle be used to determine unknown parameters.
The current experimental data however exhibit too much scatter to
determine parameters with accuracy, therefore a combined experimental
and theoretical investigation of multicellular tumor spheroids of a
single well-defined cell line is of urgent interest.

The presented model is especially suitable for systems with a
comparably large number of cells. In addition, it supports different
cell types as well. The cell shape however, is restricted to convex cells.
This makes it suitable to model rather dense cell tissues such as
e.~g.~epithelia
where one can investigate the roles of differential adhesion, 
elastic interactions and active cellular migration in tissue
flow equilibrium.
Further applications of the Voronoi/Delaunay method will therefore
include the modeling of epithelia, bone formation, and bio films.
In addition, the weighted Delaunay triangulation is a suitable tool for the modeling of boundary conditions e.~g.~in froths.


\section{Acknowledgments}\label{Sacknowledge}
G.~Schaller is indebted to T.~Beyer and W.~Lorenz for discussing many aspects of the algorithms
and testing the code. J.~Galle is being thanked for his advice on cellular interactions.
G.~Schaller was supported by the
SMWK. M.~Meyer-Hermann was supported
by a Marie Curie Intra-European Fellowship within the Sixth EU Framework Program.


\section{Appendix}\label{Sappendix}

\subsection{Program Architecture}\label{SSprogarch}

The programming language $C++$ supports object-oriented programming and thus enables us to identify individual cells with instantiations of objects. 
These objects are stored in a list to allow for efficient deletion (apoptosis or necrosis) and insertion (proliferation).
We had already implemented a weighted kinetic and dynamic Delaunay triangulation in three dimensions \cite{schaller2004} which provides -- once calculated
-- constant average access to the next neighbors for differently sized spheres. 
This is achieved by using pointers on cells as the objects in the weighted Delaunay triangulation and
storing the triangulation vertices in the cell objects.
The Voronoi tessellation -- which is the geometric dual of the Delaunay triangulation -- 
provides the three-dimensional contact surface corrections.

If the spatial steps are not too large, the neighborship can be
updated over the time with in average linear effort, 
i.~e.~the time necessary to update the neighborship relations after movement
scales linearly with the number of cells.
This limitation can be safely ensured by an adaptive step size algorithm in the numerical solution of 
equation (\ref{Eeom2}).
In our simulations, the average time step size was around $30$ s thus leading to roughly 
$60000$ time steps for $23$ days of simulation time.
At every time step the list of cells is iterated and for every cell all new variables are calculated.
Afterwards the cellular parameters are synchronized.
Note that discontinuous events such as cell proliferation and cell death correspond to insertion/deletion 
of just one cell in the list and become valid in the next time step.
The Delaunay triangulation and the diffusion grid are then updated with the cellular displacements and radius changes or nutrient consumption rates, respectively.
Therefore, all coupled equations are solved synchronously by storing the solution of every equation until the solutions of all 
equations have been calculated.

\subsection{Cellular kinetics}\label{SScellkin}

In the over-damped approximation, the cellular equation of motion
(\ref{Eeom2}) 
is just a first order differential equation that can easily 
be solved numerically.
There is a variety of established numerical algorithms to
choose from and we decided to stick with a simple forward-time
discretization -- which is just a first order method.
The first reason for this is that the uncertainties arising from the cell model presumably exceed the numerical errors by orders of magnitude.
In addition, higher order methods such as e.~g.~ Runge-Kutta require intermediate evaluations of the forces. 
In our model however this would necessitate intermediate refinements of the triangulation thus considerably increasing the numerical complexity.
Multi-value Predictor-Corrector methods are also not suitable, since
in the present model the intercellular forces are not continuous, especially during mitosis.
Keeping these arguments in mind one still has to guarantee numerical
stability of the results. This can be achieved by
using an adaptive time step size. In order to avoid slope calculations
we chose a small time step if the spatial step sizes exceeded a 
critical value, which was always chosen much smaller than the cellular
radius.

\subsection{Reaction-Diffusion Equation}

Three-dimensional reaction-diffusion equations often constitute a significant challenge for present computational hardware
since for a reasonable resolution a large number of lattice points is needed.
In addition, not every algorithm is numerically stable. For example,
the normal ADI algorithm is unconditionally stable in two dimensions
but not in three \cite{press1994}. Though there exist modified ADI
algorithms that are unconditionally stable in three dimensions as
well, the complete solution of the reaction diffusion system
(\ref{Ediffusion}) is quite intensive in three dimensions -- unless
one restricts to low resolutions. 

If the diffusion coefficients and the considered time steps are comparably large, 
the steady-state approximation $\pdiff{u}{t} \approx 0$ can be applied and
by neglecting the time dependencies equation (\ref{Ediffusion}) reduces to a Helmholtz problem
\bea\label{Ehelmholtz}
\left[\vec \nabla D(\Vektor{x})\right] \cdot \left[\vec \nabla u(\Vektor{x})\right] + D(\Vektor{x}) \vec \nabla^2 u(\Vektor{x}) = r(\Vektor{x})\,.
\eea
The steady-state-approximation has already been applied in e.~g.~\cite{groebe1996}.
Equation (\ref{Ehelmholtz}) can be solved numerically with comparably
low computational effort and -- more important -- with numerically
stable methods. Since the diffusion coefficients of both oxygen and
glucose are very large in comparison with the cellular movements, we
have decided to employ the steady-state approximation when solving the
dynamics of the nutrients.
The methods to solve (\ref{Ehelmholtz}) differ significantly in their convergence time.
A simple relaxation method such as Jacobi or Gauss-Seidel \cite{press1994} does not converge fast enough. 
In case of spatially constant diffusion coefficients the Fast Fourier Transform can be employed. Tumor tissue however, does
have a different diffusivity than agar \cite{kunz_schughart2000,freyer1986b} which made us favor a Vcycle-Multi-grid algorithm that uses 
Gauss-Seidel relaxation \cite{briggs2000}.

Since the discretization of equations (\ref{Ediffusion}) and (\ref{Ehelmholtz}) is done on a simple $64 \times 64 \times 64$ cubic lattice with a lattice
constant of 15.625 $\rm \mu m$ -- which is larger than the cellular diameter -- and as the cell positions are
arbitrary in our off-lattice model, we do use a tri-linear interpolation to determine the local concentration from the concentrations on 
the eight closest lattice nodes
\bea
  f(x,y,z) &=& f_{000} (1-x) (1-y) (1-z)\nn\\
        &&+ f_{100} x (1-y) (1-z) + f_{010} (1-x) y (1-z)\nn\\ 
        &&+ f_{001} (1-x) (1-y) z\nn\\
        &&+ f_{110} x y (1-z) + f_{101} x (1-y) z \nn\\
        &&+ f_{011} (1-x) y z + f_{111} x y z\,,
\eea
where $f_{ijk}$ represent the values of the function $f$ on the corners of a cube of length 1.
The reaction rates created by the cells are handled similarly by distributing them on the closest lattice nodes.
The local diffusion coefficients can be set by the tumor cells according to their spatial position.
This approximates the correct boundary conditions.
The size of the diffusion grid was with $1000^3$ $\rm \mu m^3$ always completely enclosing the tumor spheroids and by direct observation
of the nutrient isosurfaces it was made sure that the rectangular boundary conditions did not influence the spheroidal concentration
isosurfaces in the vicinity of the tumor spheroid.

\subsection{Fitting experimental data}

In order to minimize the difference between theoretical and experimental observables we performed roughly $150$ computer simulations 
over a wide range of parameters until the visual agreement with the experiment was satisfactory.
Afterwards we started Powells method \cite{press1994} with several perturbations around this optimal parameter set
by minimizing the squared differences of the logarithms of theoretical and experimental growth curves, i.~e.
\bea
\chi^2 = \sum_{i \rm:exp} \sum_{j \rm:meas} 
        \frac{1}{\sigma_{ij}^2} \left\{\ln N_{ij}^{\rm exp} - \ln N_{ij}^{\rm sim}[p_1, p_2, \ldots]\right\}^2\,,
\eea
where the $p_\alpha$ are the parameters that have been varied and the errors of the experimental data points $\sigma_{ij}$ have been estimated by calculating
the difference to a Gompertz growth curve.
Note that it is a purely geometric and therefore deterministic
algorithm, which opens the possibility that it will terminate within a
local minimum.
In order to decrease the probability of terminating within a local
minimum, several runs should be performed.
However, the changes of parameters are negligible, since
due to the strong scatter of the data the visual data fit is
satisfactory already.


\end{document}